\documentclass{aastex631}

\begin{document}

\title{Analysis of Multi-Wavelength Data for 30 Galactic Emission-Line Stars with Dusty Envelopes.}

\correspondingauthor{Nazgul Alimgazinova}
\email{Nazgul.Alimgazinova@kaznu.kz}

\author[0000-0001-6653-2948]{A.~Zh.~Naurzbayeva}
\affiliation{Al-Farabi Kazakh National University, Al-Farabi Ave., 71, 050040, Almaty, Kazakhstan}

\author[0000-0002-3833-1038]{A.~S.~Miroshnichenko}
\affiliation{Department of Physics and Astronomy, University of North Carolina at Greensboro, P.O. Box 26170, Greensboro, NC 27402-6170, USA}
\affiliation{Fesenkov Astrophysical Institute, Observatory, 23, Almaty, 050020, Kazakhstan}
\affiliation{Al-Farabi Kazakh National University, Al-Farabi Ave., 71, 050040, Almaty, Kazakhstan}

\author[0000-0002-4596-1855]{N.~Sh.~Alimgazinova}
\affiliation{Al-Farabi Kazakh National University, Al-Farabi Ave., 71, 050040, Almaty, Kazakhstan}

\author[0000-0003-2526-2683]{S.~V.~Zharikov}
\affil{Universidad Nacional Aut\'{o}noma de M\'{e}xico, Instituto de Astronom\'{i}a, AP 106,  Ensenada 22800, BC, M\'{e}xico}
\affil{Al-Farabi Kazakh National University, Al-Farabi Ave., 71, 050040, Almaty, Kazakhstan}

\author[0000-0001-7145-9888]{C.~Rossi}
\affiliation{Dipartimento di Fisica, Universit\'a Sapienza, Piazzale Aldo Moro 2, I 00185 Roma, Italy}

\author[0000-0001-9944-8398]{I.~V.~Reva}
\affiliation{Fesenkov Astrophysical Institute, Observatory, 23, Almaty, 050020, Kazakhstan}

\author[0000-0002-5973-5074]{R.~I.~Kokumbayeva}
\affiliation{Fesenkov Astrophysical Institute, Observatory, 23, Almaty, 050020, Kazakhstan}

\author[0000-0002-1672-894X]{Ch.~T.~Omarov}
\affiliation{Fesenkov Astrophysical Institute, Observatory, 23, Almaty, 050020, Kazakhstan}

\author[0000-0001-5981-2255]{A.~Raj}
\affil{Indian Centre for Space Physics, 466 Barakhola, Netai Nagar, Kolkata 700099, India}

\author{M.~S.~Bisht}
\affil{Indian Centre for Space Physics, 466 Barakhola, Netai Nagar, Kolkata 700099, India}

\author[0000-0002-8988-8434]{A.~Joshi}
\affil{Institute of Astrophysics, Pontificia Universidad Cat\'olica de Chile, Av. Vicu\~na MacKenna 4860, 7820436 Santiago, Chile}

\author[0000-0001-7920-5372]{A.~K.~Kuratova}
\affiliation{Al-Farabi Kazakh National University, Al-Farabi Ave., 71, 050040, Almaty, Kazakhstan}

\author[0000-0003-3096-759X]{R.~J.~Rudy}
\affiliation{Kookoosint Scientific, 1530 Calle Portada, Camarillo, CA 93010, USA}

\author[0000-0002-2291-655X]{R.~C.~Puetter}
\affiliation{Center for Astrophysics \& Space Science University of California, San Diego, 9500 Gilman Dr., La Jolla, CA 92093, USA}

\author[0000-0002-3389-1722]{R.~B.~Perry}
\affiliation{Alabaster Scientific, P.O. Box 120, Irvington, VA 22480, USA}

\author[0000-0001-7548-0190]{N.~Manset}
\affil{Canada-France-Hawaii Telescope Corporation, 65-1238 Mamalahoa Hwy, Kamuela, HI 96743, USA}

\author[0000-0002-7449-0108]{N.~L.~Vaidman}
\affil{Al-Farabi Kazakh National University, Al-Farabi Ave., 71, 050040, Almaty, Kazakhstan}
\affiliation{Fesenkov Astrophysical Institute, Observatory, 23, Almaty, 050020, Kazakhstan}

\author[0009-0005-6851-7270]{Sh.~T.~Nurmakhametova}
\affil{Al-Farabi Kazakh National University, Al-Farabi Ave., 71, 050040, Almaty, Kazakhstan}

\author[0000-0002-0322-1509]{A.~B.~Manapbayeva}
\affil{Kazakh National Women’s Teacher Training University.	Aiteke Bi, 99, 050000, Almaty, Kazakhstan}

\author[0000-0001-5163-508X]{S.~A.~Khokhlov}
\affil{Al-Farabi Kazakh National University, Al-Farabi Ave., 71, 050040, Almaty, Kazakhstan}



\begin{abstract}

We selected 30 stars with the $V$--band brightness of 11--16 mag, visual color-indices $B-V \lesssim 1.3$ mag, and large near-IR color-indices ($J-H$ and $H-K \gtrsim 0.7$ mag) to search for the presence of the B[e] phenomenon.
Spectroscopic observations confirmed the presence of emission lines previously detected in 11 stars of the sample and revealed them in 4 other stars for the first time.
Spectral energy distributions (SED) in the optical and IR spectral regions were analyzed and compared with theoretical SEDs. The fundamental parameters were estimated using the SED fits corrected for the interstellar and circumstellar (CS) extinction and {\it Gaia} DR3 distance measurements. Properties of 14 objects previously classified as pre-main-sequence stars were reevaluated.
The B[e] phenomenon was found in 16 objects, most of which exhibit observed properties of FS\,CMa type objects, a still small group of Galactic stars or binary systems with predominantly hot CS dust. Properties of 2 objects (IRAS\,06237+0020 and IRAS\,19156$-$0935) are consistent with those of Proto-Planetary Nebulae at early phases of this evolutionary stage.
The nature and evolutionary status of 11 sample objects remains uncertain and need further investigation.
\end{abstract}



\section{Introduction} \label{sec:intro}

Stars or stellar systems, whose radiation deviates from that is expected for an undisturbed photosphere and atmosphere, are important laboratories for understanding non-stationary processes caused by evolutionary changes and/or interactions between the system's components. Such deviations include the presence of line emission, excesses of optical and infrared (IR) radiation, deficits of UV radiation, brightness, and spectral variations. Examples of hot (OB-type) stars that exhibit such phenomena include classical Be stars, objects with the B[e] phenomenon, Wolf-Rayet stars, Herbig Ae/Be stars, etc. Detecting stars surrounded by CS matter and constraining their properties using various observational methods (spectroscopy, photometry, polarimetry, interferometry) allows us to closely study processes that occur at different evolutionary stages and thus get better insights into the evolution of stars and stellar systems.

Emission lines in the spectra of stars manifest the presence of circumstellar (CS) material that contains ionized gas. The latter also produces an excess of continuum radiation in the IR and even the optical spectral region, due to free-free radiation from electrons ionized by stellar UV photons. Additionally, a large IR excess indicates the presence of CS dust located at distances from the star larger than those of the ionized gas due to the low dust sublimation temperature ($\le$ 1,500--2,000 K). The existence of CS matter is typically due to stellar winds of evolved massive and luminous stars, a strong mass-loss at late stages of the evolution of low- and intermediate-mass stars (e.g., at the AGB and post-AGB stage),  or mass-transfer between the components in binary or multiple systems.

Several sky surveys aimed at finding emission-line stars have been conducted in the optical spectral range over the last $\sim$100 years. They include the Mount Wilson observatory survey \citep[MWC and AS,][]{1933ApJ....78...87M,1950ApJ...112...72M}, the Hamburg-Bergedorf survey \citep{1999A&AS..134..255K}, and LAMOST survey \citep{2022ApJS..259...38Z}, to name a few.
These surveys were purely spectroscopic and thus provided only limited information about the discovered emission-line objects. This information typically included spectral types, emission-line strengths and content, but was insufficient for the evolutionary classification that requires more constraints from other methods (e.g., multicolor photometry, direct imaging, etc.).

Purely photometric surveys, such as ASAS \citep{1997AcA....47..467P,2017PASP..129j4502K} in the optical region and IRAS \citep{1988iras....7.....H} or WISE \citep{2013wise.rept....1C} in the IR region, especially combined to cover a wide spectral range, are capable of revealing the excess of IR radiation. However, these data may be insufficient for constraining the star's fundamental parameters due to similar effects of the interstellar extinction and CS envelope on its SED.

In order to find new objects with CS envelopes and/or constrain the properties of already discovered but poorly studied ones, we undertook a long-term search for those with specific IR color-indices, completed these with optical photometry to construct the SEDs, and took spectra of a sample of most promising candidates for early-type stars with gaseous-and-dusty envelopes. Our main goal was to find more objects to a still small group of those with the B[e] phenomenon, whose main features are permitted and forbidden emission lines and large IR excesses \citep{1998A&A...340..117L,2007ApJ...667..497M}. This group represents objects which undergo very different evolutionary stages (from the pre-main-sequence to very evolved supergiants and proto-planetary nebulae) but have similar conditions in their CS envelopes. However, the mechanisms of the CS matter formation are not always clear, and a number of objects with this phenomenon have uncertain classification.

The results of our previous searches were published by \citet{2007ApJ...671..828M,2017ASPC..508..229K,2025Galax..13...67N}. We found over 30 objects with the B[e] phenomenon, including a dozen pre-main-sequence star candidates, and a few emission-line candidates of other possible nature (e.g., Be stars). Our team published more detailed studies of some of them in separate papers, such as \citet{2015ApJ...809..129M,2018ApJ...856..158K,2022ApJ...932...36K}.

In this paper, we present our findings for 30 objects, four of which were mentioned by \citet{2017ASPC..508..229K} without details (see Sect.\,\ref{properties}). The search criteria and strategy are briefly discussed in Sect.\,\ref{search}, observations are described in Sect.\,\ref{observ}, properties of individual objects are summarized in Sect.\,\ref{properties}, a discussion of the findings is provided in Sect.\,\ref{discussion}, and a summary of the project is given in Sect.\,\ref{conclusions}.

\section{Search for hot stars with IR excesses} \label{search}

As our main goal was finding hot stars with relatively warm CS dust, a feature of some objects with the B[e] phenomenon, such as FS\,CMa systems
\citep[c.f.][]{2007ApJ...667..497M,2007ApJ...671..828M}, the main selection criteria were near-IR color-indices, namely $J-H$ and $H-K$. In particular, we only selected objects with both $J-H$ and $H-K \gtrsim$ 0.7 mag. This constraint came from our previous studies of objects with the B[e] phenomenon \citep[see Figure 2 in][]{2017ASPC..508..229K}. 
In order to cut off late-type stars, we limited the search to objects with optical color-indices of $B-V \leq 1.3$ mag. This criterion is not very strict and does not pick up heavily reddened hot stars with color-excesses of E$(B-V) \geq 1.5$ mag, but it reduces the number of candidates for further investigation. The latter condition is important in view of the limited amount of telescope time available to us and relative faintness of the targets. This sample is also limited to objects that can be observed from latitudes of $\geq 20^{\circ}$ in the Northern hemisphere, where the telescopes used for this project were available to us.

A convenient catalog for our search was initially NOMAD \citep{2004AAS...205.4815Z} and later UCAC4 \citep{2013AJ....145...44Z}, where some optical \citep[from APASS,][]{2009AAS...21440702H} and near-IR \citep[from 2MASS,][]{2006AJ....131.1163S} magnitudes along with accurate astrometric positions were collected.
Photometric information for the objects selected with the above procedure was supplemented by longer-wavelength IR fluxes from WISE \citep{2013wise.rept....1C}, AKARI \citep{2010A&A...514A...1I}, MSX \citep{2003yCat.5114....0E}, IRAS \citep{1988iras....7.....H}, and Herschel \citep{2020yCat.8106....0H} catalogs to constrain the shape of the IR excess. Most of the photometric data for the selected objects are listed in Table\,\ref{tab1}.

\begin{figure*}[t]
 \centering
    \includegraphics[width=1.\linewidth]{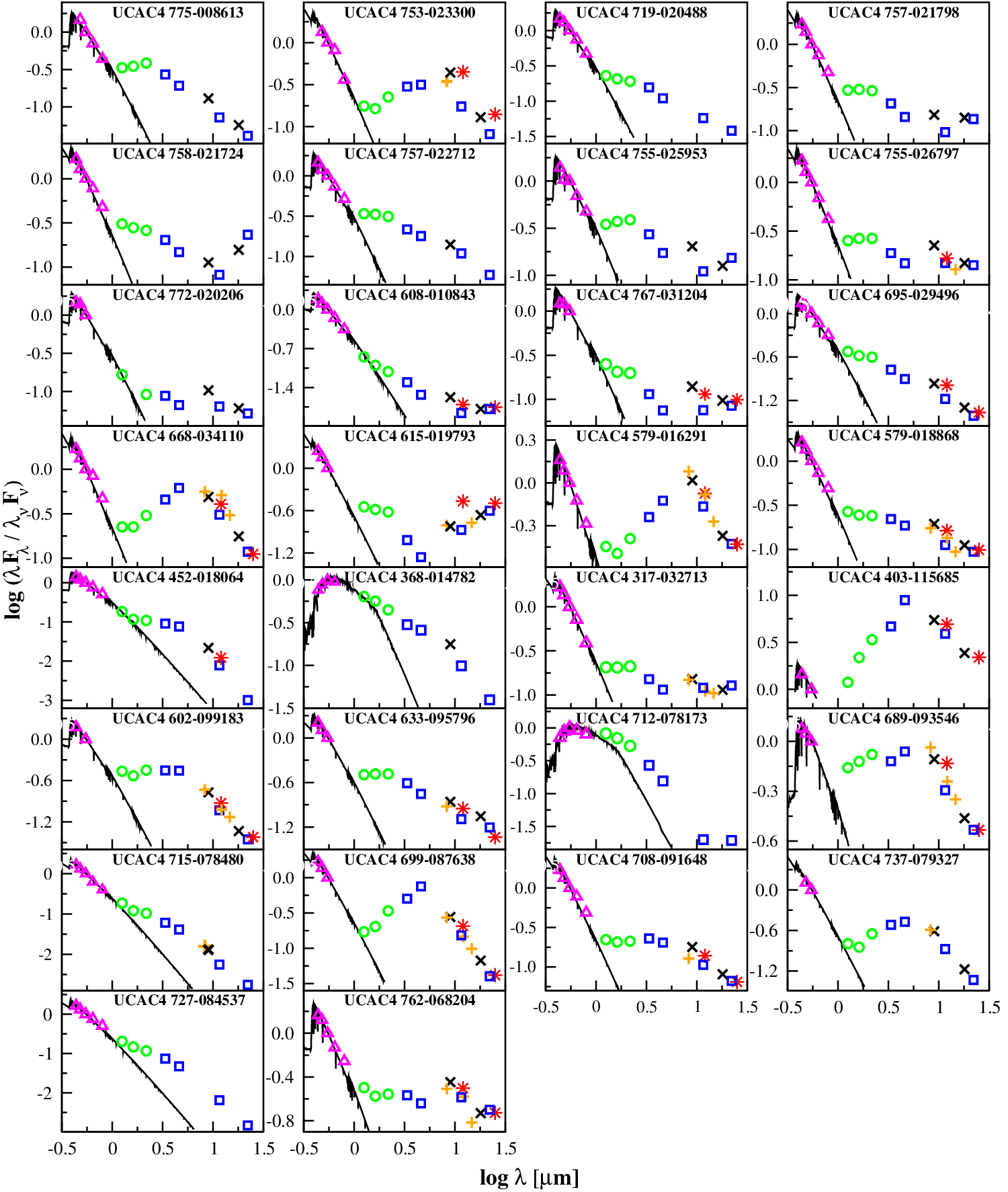} 
\caption{Objects' dereddened SEDs. Fluxes are normalized to that in the $V$--band. Symbols: triangles -- optical photometry, green circles -- 2MASS $JHK$ data, blue squares -- WISE data, orange pluses -- MSX data, black crosses -- AKARI 9--$\mu$m \& 18--$\mu$m data, and red asterisks -- IRAS 12 \& 25--$\mu$m data). Best-fitting models are shown by the solid lines. }
\label{fig1}
\end{figure*}

The SEDs shown in Fig.\,\ref{fig1} were constructed from available photometric data and corrected for the total (interstellar + possible CS) extinction, assuming that both extinction components follow the same average Galactic interstellar extinction law \citep{1979ARA&A..17...73S}. The reddening corrections were determined by fitting the observed SEDs in the optical range ($\lambda \le 1 \mu$m) to the theoretical models of \citet{2003IAUS..210P.A20C} using the procedure described in \citet{2026Galax..14...32N}. The latter compared the observed SEDs composed of broadband photometric data with synthetic data calculated from the theoretical SEDs convolved with the response curves of the filters available in the literature. This procedure is not extremely precise for deriving the objects' effective temperatures (T$_{\rm eff}$), but it is consistent with the spectral types estimated from the available spectra, which were taken at several telescopes in a wide range of spectral resolving powers (see Sect.\,\ref{observ}). The accuracy of the derived T$_{\rm eff}$ can be estimated as $\sim \pm2,000$ K (1--2 spectral subtypes) for most objects {\bf that corresponds to the accuracy of the derived color-excesses E$(B-V) of \sim0.05$ mag}. 
Only for 3 objects with T$_{\rm eff} \leq 10,000$~K (see Table\,\ref{tab2}), the uncertainty of T$_{\rm eff}$ is smaller.

As all objects show IR excesses due to CS material, only a few data points (typically $BVgr^{\prime}i^{\prime}$ photometry from the UCAC4 catalog) were available in the optical part of the SED to find the best combination of the color-excess and T$_{\rm eff}$ to represent the radiation of the underlying star. The procedure is not sensitive to surface gravity of the star, but it is backed up by an analysis of the spectroscopic data that include our optical and near-IR spectra and {\it Gaia} low-resolution spectra.
The BP/RP {\it Gaia} spectra were dereddened using the color-excesses derived from the SED fitting and compared with dereddened SEDS of the objects. In particular, the Balmer jumps detected in the {\it Gaia} spectra were compared to those of the best-fitting theoretical SEDs.

The procedure described above resulted in finding 30 objects, which can be characterized as stars (mostly early-type) with large IR excesses. All the objects exhibit emission lines in their spectra as a result of the presence of ionized gas in the CS environments. Although CS gas contributes to the IR excess, its strength and shape can only be explained by radiation from CS dust.  
IR fluxes of 19 objects were previously detected by the IRAS mission, and the H$\alpha$ lines in emission were previously detected in 19 objects (13 of them are IRAS sources) in the Hamburg--Bergedorf objective prism survey \citep{1999A&AS..134..255K}.
The results of the SED fitting procedure that include the objects' T$_{\rm eff}$, color-excesses E$(B-V)$, luminosity calculated using {\it Gaia} DR3 distances \citep{2021AJ....161..147B} and bolometric corrections, which correspond to the derived T$_{\rm eff}$, are presented in Table\,\ref{tab2}.

\begin{deluxetable*}{lccccccrcccccrl}[htb]
\tabletypesize{\scriptsize}
\tablewidth{0pt} 
\tablecaption{List of objects with IR excesses selected using photometric criteria for those with the B[e] phenomenon \label{tab1}}
\tablehead{
\colhead{UCAC4} & \colhead{IRAS} & \colhead{R.A.(2000)} & \colhead{Decl.(2000)} & \colhead{$V$} & \colhead{$B-V$} & \colhead{$J-K$} & \colhead{$K$} & \colhead{$F_{12}$} & \colhead{$F_{25}$} & \colhead{$W_{1}$} & \colhead{$W_{2}$} & \colhead{$W_{3}$} & \colhead{$W_{4}$} & \colhead{Sp.T.} \\
\colhead{} & \colhead{} & \colhead{h m s} & \colhead{d m s} & \colhead{mag} & \colhead{mag} & \colhead{mag} & \colhead{mag} & \colhead{Jy} & \colhead{Jy} & \colhead{mag} & \colhead{mag} & \colhead{mag} & \colhead{mag} & \colhead{}
}
\startdata
\hline
1 & 2 & 3 & 4 & 5 & 6 & 7 & 8 & 9 & 10 & 11 & 12 & 13 & 14 & 15 \\
\hline
775-008613 &		    & 01 39 32.6 & +64 53 02.3 & 14.57 & $-$  & 2.32 & 9.11 & $-$ &  $-$  & 7.95 & 7.26 & 5.41 & 3.90 &	   \\
753-023300 & 01571+6018 & 02 00 39.5 & +60 32 59.1 & 15.51 & $-$  & 2.57 &	9.88 & 1.17 & 0.81 & 7.96 & 6.79 & 4.57 & 3.23 & B    \\
719-020488 & 02162+5326 & 02 19 35.1 & +53 40 19.8 & 13.64 & 0.38 &	1.64 & 10.81 & $-$	& $-$  & 9.63 & 9.02 & 6.80 & 5.16 &	   \\
757-021798 & 02258+6102 & 02 29 35.9 & +61 15 56.9 & 14.36 & 1.02 &	2.20 &	8.94 & 0.63 & 1.32 & 7.09 & 7.08 & 4.60 & 2.08 &	   \\
758-021724 &	        & 02 34 10.3 & +61 24 40.4 & 14.23 & 0.83 &	1.92 &	9.51 &	$-$	& $-$  & 8.16 & 7.50 & 5.31 & 2.12 & B8 {\sc v}e \\
757-022712 &		    & 02 39 50.9 & +61 18 28.9 & 14.37 & 0.70 &	1.90 & 10.19 &	$-$ & $-$  & 9.13 & 8.31 & 5.93 & 4.49 & A2e	\\
755-025953 &	        & 02 47 48.6 & +60 57 50.5 & 14.32 & 0.73 &	2.08 &	9.96 &	$-$ & $-$  & 8.98 & 8.44 & 5.93 & 3.48 & A0--1{\sc v}e	\\
755-026797 & 02551+6042 & 02 59 05.1 & +60 54 04.1 & 13.31 & 0.79 &	2.15 &	8.67 & 1.00 & $-$  & 7.54 & 6.75 & 3.83 & 1.77	& B4 {\sc v}e \\
772-020206 &		    & 03 09 43.5 & +64 19 04.5 & 12.94 & 0.76 &	1.36 &	9.95 & $-$  & $-$  & 8.51 & 7.78 & 4.91 & 3.03	&		\\
608-010843 & 03519+3123 & 03 55 03.8 & +31 31 57.3 & 12.14 & 0.87 & 1.43 &	8.86 & 0.34 & 0.66 & 7.84 & 7.35 & 5.26 & 2.96	&		\\
767-031204 & 04141+6307 & 04 18 41.2 & +63 14 40.7 & 13.99 & 1.04 &	1.80 &	9.90 & 0.29 & 0.54 & 9.11 & 8.42 & 5.51 & 3.27	&	    \\
695-029496 & 04265+4845 & 04 30 16.2 & +48 52 10.0 & 13.27 & 0.92 &	1.86 &	8.88 & 0.53 & 0.48 & 7.82 & 7.10 & 4.87 & 3.34	&       \\
668-034110 & 04577+4328 & 05 01 20.3 & +43 32 50.8 & 12.55 & 0.35 &	2.21 &	8.91 & 1.53 & 0.89 & 7.04 & 5.71 & 3.54 & 2.49	&       \\
615-019793 & 05111+3244 & 05 14 26.9 & +32 48 03.3 & 12.25 & 0.46 &	1.77 &	8.48 & 2.51 & 5.03 & 7.91 & 7.50 & 3.73 & 0.95	&       \\
579-016291 & 05311+2534 & 05 34 17.0 & +25 36 16.3 & 13.09 & 0.38 &	1.96 &	9.49 & 1.34 & 1.25 & 7.71 & 6.43 & 3.61 & 2.18	&  B2   \\
579-018868 & 05481+2541 & 05 51 17.5 & +25 42 30.0 & 12.82 & 0.87 &	2.00 &	8.15 & 1.77 & 2.32 & 6.73 & 5.86 & 3.50 & 1.57	&  Be   \\
452-018064 & 06237+0020 & 06 26 20.0 & +00 18 44.2 & 11.19 & 1.03 &	1.55 &	7.40 & 0.59 & $-$  & 6.08 & 5.21 & 4.77 & 4.87	&  M2--3 \\
368-014782 &		    & 06 30 16.5 &$-$16 29 37.5& 13.99 & 1.04 &	1.42 & 10.38 &	$-$	& $-$  & 9.42 & 8.59 & 6.71 & 5.62	&        \\
317-032713 &            & 07 48 39.0 &$-$26 40 40.3& 13.37 & 0.61 & 2.02 &	9.41 &	$-$	& $-$  & 8.31 & 7.58 & 4.61 & 2.44	&        \\
403-115685 & 19156--0935& 19 18 24.0 &$-$09 30 15.0& 11.94 & 0.40 &	2.96 &	6.00 &23.50 & 22.20& 4.08 & 3.04 & 0.59 &$-$0.88&		 \\
602-099183c& 19552+3005 & 19 57 12.4 & +30 13 16.0 & 13.45 & 1.06 &	2.20 &	8.10 & 0.93 & 0.65 & 6.58 & 5.52 & 4.05 & 2.98	&        \\
633-095796 & 20098+3622 & 20 11 43.5 & +36 31 08.5 & 13.05 & 0.50 &	1.95 &	9.08 & 0.34 & 0.30 & 7.95 & 7.30 & 5.23 & 3.41	&        \\
712-078173 &            & 21 01 44.1 & +52 17 17.7 & 13.25 & 1.22 &	1.38 &	9.18 & $-$  & $-$  & 8.50 & 8.10 & 7.41 & 5.31	&        \\
689-093546 & 21095+4726 & 21 11 19.1 & +47 38 47.6 & 12.45 & 0.61 &	2.03 &	8.03 & 2.19 & 1.85 & 6.73 & 5.59 & 3.25 & 1.76	&        \\
715-078480 &            & 21 27 26.9 & +52 59 25.7 & 11.15 & 0.71 & 1.42 &  7.77 &  $-$ &  $-$ & 6.86 & 5.24 & 5.50 & 4.64  &  OBe   \\
699-087638 & 21263+4927 & 21 28 05.4 & +49 41 00.2 & 12.60 & 0.79 &	2.83 &	7.69 & 2.40 & 1.05 & 5.75 & 4.27 & 3.09 & 2.42	&        \\
708-091648 & 21525+5113 & 21 54 21.2 & +51 27 53.6 & 12.84 & 0.60 &	1.96 &	8.92 & 0.80 & 0.80 & 7.35 & 6.45 & 4.24 & 2.64	&  Be     \\
737-079327 &		    & 21 56 28.5 & +57 14 45.5 & 15.89 & 0.41 &	2.65 & 10.31 & $-$  & $-$  & 8.39 & 7.18 & 5.29 & 4.29  &         \\
727-084537 &		    & 22 21 11.0 & +55 15 46.2 & 12.14 & 0.30 &	1.24 &	9.78 & $-$	& $-$  & 8.86 & 8.35 & 7.58 & 7.10	&  B0     \\
762-068204 & 22541+6205 & 22 56 07.7 & +62 21 48.2 & 13.51 & 1.08 &	2.01 &	8.43 & 2.35 & 3.05 & 6.92 & 6.04 & 2.99 & 1.15	&         \\
\enddata
\tablecomments{
Column information: 
(1) -- ID from UCAC4, 
(2) -- ID from IRAS, 
(3) -- Right Ascension in J2000, 
(4) -- Declination in J2000, 
(5) -- $V$--band magnitude, 
(6) -- $B-V$ color-index from UCAC4, 
(7) -- $J-K$ color-index from 2MASS, 
(8) -- $K$--band magnitude from 2MASS, 
(9) -- flux density in the IRAS 12~$\mu$m band in Janskys, 
(10) -- flux density in the IRAS 25~$\mu$m band in Janskys,
(11) -- W1-band (3.35~$\mu$m) magnitude from WISE, 
(12) -- W2-band (4.6~$\mu$m) magnitude from WISE, 
(13) -- W3-band (11.6~$\mu$m) magnitude from WISE, 
(14) -- W4-band (22.1~$\mu$m) magnitude from WISE, 
(15) -- spectral type listed in the SIMBAD database.
}
\end{deluxetable*}

\section{Observations} \label{observ}

Our spectroscopic observations were obtained in 2010--2026 with the telescopes and spectrographs listed below. Low-resolution optical and near-IR spectra (spectral resolving power $R = \lambda/\Delta\lambda = 800-1500$) were taken at the following facilities: 1.52\,m telescope of the Bologna Observatory (Mt. Orzale, Italy, Bologna Faint Object Spectrograph, BFOSC), 2\,m Himalayan Chandra Telescope (HCT, Hanle, India, Himalayan Faint Object Spectrograph Camera, HFOSC), 2.12\,m telescope of the Observatorio Astron\'omico Nacional San Pedro Martir Observatory (OAN SPM, Mexico, Boller-Chivens spectrograph), and the 3\,m telescope of the Lick Observatory (CA, USA, 0.46--2.5 $\mu$m, NIRIS spectrograph). The spectra taken at Lick were flux-calibrated using various standard stars, which were observed immediately after the science target at nearly the same airmass. The typical procedure for calculating the spectrum in energy units is outlined in \citet{1999AJ....118..666R}.

Medium-resolution optical spectra were taken with the REOSC \'echelle spectrograph on the 2.12\,m telescope of OAN SPM ($R \sim 18,000$; we refer to this telescope as just SPM hereafter). High-resolution spectra were taken on the 3.6\,m Canada-France-Hawai'i Telescope (CFHT, Maunakea, HI, USA) with the ESPaDoNS spectropolarimeter ($\lambda\lambda 0.36-1.05 \mu$m, $R \sim 65,000$) and on the 2.7\,m Harlan J. Smith telescope of the McDonald Observatory (Mt. Locke, TX, USA, Tull spectrograph, $\lambda\lambda 0.36-1.05 \mu$m, $R \sim 60,000$).

Data taken at the above mentioned telescopes except for the CFHT data were reduced with standard techniques (bias subtraction, flat-fielding, and wavelength calibration with various calibration lamps filled with ThAr for higher-resolution instruments and other gasses for lower-resolution instruments). The CFHT data were reduced with Upena and Libre-ESpRIT software \citep{1997MNRAS.291..658D}.

Photometric observations of 8 objects from the studied group were obtained at a 1\,m telescope of the Tien-Shan Astronomical Observatory (TShAO) of the Fesenkov Astrophysical Institute (near Almaty, Kazakhstan). A $3056 \times 3056$ Apogee F9000 D9 CCD camera with 12--$\mu$m pixels and a 3--step cooling was used with a set of $BVR_{\rm c}$ filters. The data reduction process was standard and was briefly described by \citet{2025Galax..13...67N}.
Comparison stars were chosen from images of the objects' areas using photometry from the UCAC4 catalog and checking for variability using their light curves from the All-Sky Automated Survey for Supernovae survey \citep[ASAS SN,][]{2017PASP..129j4502K}. Transformations from the TShAO instrumental photometric system into the Johnson-Cousins system were performed using images of standard fields (e.g., the open cluster NGC\,2169). 

The purpose of the photometric observations was to enlarge the number of multicolor data and extend their temporal coverage, which is very useful for strongly variable objects. 
The TShAO photometric data are shown in Table\,\ref{tab3} and agree well with the ASAS SN data for all the observed objects. The measurement uncertainties range from 0.01 to 0.03 mag.

\begin{table*}
\caption{Parameters of the studied objects}\label{tab2}
    \setlength{\tabcolsep}{2pt}
\begin{center}
\begin{tabular}{crcrcclcl}
\hline\noalign{\smallskip} 
 UCAC4 ID & $T_{\rm eff}$        & $\log(L/L_{\odot})$ & EW(H$\alpha$)          & $E(B\!-\!V)$ & $E(g\!-\!r)$ & Dist. & Slope & Obs. \\
          & \multicolumn{1}{c}{K}&                     & \multicolumn{1}{c}{\AA}& mag          &  mag         & pc    &       &      \\
\hline
1 & 2 & 3 & 4 & 5 & 6 & 7 & 8 & 9 \\
\hline
775-008613	&	10,000	&	$ 1.38  \pm	0.32$	&	--30.3	&	1.1	&	0.4	&	$ 811	_{-	13	 }	^	{+	19	 }$	&	-1.0	& SPM	\\ 
753-023300	&	21,000	&	$ 3.23	\pm 0.37$	&	--275.9	&	1.4	&	1.2	&	$ 3063	_{-	268	 }	^	{+	281	 }$	&	-1.3	& MO	\\
719-020488	&	11,000	&	$ 2.11	\pm 0.28$	&	--9.1	&	0.5	&	0.2	&	$ 2952	_{-	138	 }	^	{+	147	 }$	&	-0.7	& IAO	\\
757-021798	&	21,000	&	$ 3.32	\pm 0.33$	&	--78.8	&	1.2	&	0.9	&	$ 2575	_{-	100	 }	^	{+	110	 }$	&	-0.6	& MO	\\
758-021724	&	19,000	&	$ 2.79	\pm 0.32$	&	--103.0	&	1.0	&	0.7	&	$ 2011	_{-	73	 }	^	{+	63	 }$	&	-0.7	& MO	\\
757-022712	&	10,500	&	$ 1.90	\pm 0.31$	&	--38.2	&	0.8	&	0.7	&	$ 2175	_{-	78	 }	^	{+	101	 }$	&	-0.7	& SPM	\\
755-025953	&	10,000	&	$ 1.80	\pm 0.30$	&	--31.6	&	0.7	&	1.3	&	$ 2087	_{-	65	 }	^	{+	86	 }$	&	-0.6	& MO  \\	
755-026797	&	19,000	&	$ 3.20	\pm 0.30$	&	--36.5	&	1.0	&	0.7	&	$ 2224	_{-	107	 }	^	{+	105	 }$	&	-0.2	& SPM	\\
772-020206	&	11,000	&	$ 2.04	\pm 0.28$	&	--31.9	&	0.8	&	0.6	&	$ 1178	_{-	41	 }	^	{+	48	 }$	&	-0.2	& SPM	\\
608-010843	&	12,000	&	$ 1.66	\pm 0.27$	&	--20.3	&	1.0	&	0.2	&	$ 362	_{-	2	 }	^	{+	3	 }$	&	-0.9	& SPM, MO	\\
767-031204	&	10,000	&	$ 1.53	\pm 0.30$	&	--5.67	&	0.9	&	0.6	&	$ 1042	_{-	19	 }	^	{+	18	 }$	&	-0.3	& LO	\\
695-029496	&	10,000	&	$ 2.48	\pm 0.29$	&	--14.2	&	0.9	&	0.7	&	$ 2170	_{-	79	 }	^	{+	78	 }$	&	-0.7	& MO, LO  \\	
668-034110	&	21,000	&	$ 3.28	\pm 0.26$	&	--193.3	&	0.6	&	0.3	&	$ 2799	_{-	72	 }	^	{+	78	 }$	&	-1.6	& SPM, LO	\\
615-019793	&	25,000	&	$ 3.41	\pm 0.27$	&	--77.2	&	0.7	&	0.7	&	$ 1893	_{-	80	 }	^	{+	92	 }$	&	-2.0	& CFHT, SPM, LO\\	
579-016291	&	10,500	&	$ 1.03	\pm 0.27$	&	--9.7	&	0.4	&	0.4	&	$ 701	_{-	11	 }	^	{+	10	 }$	&	-1.1	&	SPM, LO	\\
579-018868	&	16,000	&	$ 3.11	\pm 0.29$	&	--39.7	&	1.0	&	0.3	&	$ 1697	_{-	49	 }	^	{+	62	 }$	&	-0.4	&	MO  \\	
452-018064	&	10,000	&	$ 3.93	\pm 0.28$	&	--0.7   &	1.0	&	0.4	&	$ 3728	_{-	388	 }	^	{+	418	 }$	&	-2.3	&	SPM	\\
368-014782	&	5,500	&	$ 1.85	\pm 0.29$	&	--13.6	&	0.4	&   0.2	&	$ 3150	_{-	142	 }	^	{+	202	 }$	&	-1.0	&	IAO, LO \\	
317-032713	&	18,000	&	$ 3.30	\pm 0.32$	&	--51.4	&	0.8	&	0.2	&	$ 3156	_{-	374	 }	^	{+	565	 }$	&	-0.2	&	MO  \\	
403-115685	&	10,500	&	$ 1.23	\pm 0.25$	&	--3.9	&	0.5	&	0.2	&	$ 507	_{-	27	 }	^	{+	30	 }$	&	-1.0	&	MO, SPM, LO \\	
602-099183	&	11,000	&	$ 4.06	\pm 0.32$	&	--53.2	&	1.1	&	1.0	&	$ 9667	_{-	1136 }	^	{+	919	 }$ &	-1.5	&   MO, LO  \\	
633-095796	&	15,000	&	$ 3.63	\pm 0.36$	&	--27.4	&	0.7	&	1.1	&	$ 6333	_{-	1212 }	^	{+	1659 }$ &	-0.7	&  SPM, LO  \\
712-078173	&	5,500	&	$ 0.76	\pm 0.27$	&	--12.2	&	0.5	&	1.4	&	$ 557	_{-	3	 }	^	{+	4	 }$	&	-1.8    &  IAO, LO  \\	
689-093546	&	8,000	&	$ 2.23	\pm 0.26$	&	--44.3	&	0.5	&	0.5	&	$ 2390	_{-	53	 }	^	{+	63	 }$	&	-1.1    &  SPM, CFHT, LO \\
715-078480  &   16,000  &   $ 3.52	\pm 0.25$   &   --31.9  &   0.9 &   0.7 &   $ 1533  _{-28    }   ^  {+  29   }$ &   -1.8    &  CFHT, SPM, LO  \\    
699-087638	&	19,000	&	$ 3.54	\pm 0.28$	&	--43.6	&	1.0	&	0.6	&	$ 2361	_{-	44	 }	^	{+	53	 }$	&	-1.9    &  CFHT, SPM, LO \\
708-091648	&	21,000	&	$ 3.89	\pm 0.29$	&	--88.0	&	0.8	&	0.4	&	$ 4500	_{-	190	 }	^	{+	351	 }$	&	-0.7	&  CFHT, SPM, LO \\	
737-079327	&	25,000	&	$ 4.08	\pm 0.62$	&	--174.3	&	1.4	&	1.0	&	$ 8162	_{-	2201 }	^	{+	4689 }$	&	-1.8	&  MO	         \\
727-084537	&	17,000	&	$ 3.44	\pm 0.26$	&	--47.3	&	0.5	&	0.4	&	$ 3923	_{-	173	 }	^	{+	193	 }$	&	-1.9	&  MCD, SPM      \\
762-068204	&	10,500	&	$ 1.95	\pm 0.31$	&	--25.5	&	1.1	&	1.1	&	$ 903	_{-	34	 }	^	{+	40	 }$	&	-0.6	&  SPM	 \\
   \noalign{\smallskip}\hline
\end{tabular}
\end{center}
\tablecomments{Column information: 
(1) -- ID from the {\it UCAC4} catalog,
(2) -- star's effective temperature in degrees Kelvin,
(3) -- logarithm of the luminosity in solar units, 
(4) -- Equivalent width of the H$\alpha$ emission line in Angstr\'oms, 
(5) -- color-excess E($B-V$) from the SED fitting,  
(6) -- color-excess E($g-r$) at the {\it {\it {\it {\it Gaia}}}} DR3 distance from the Galactic dust maps by \citet{2019ApJ...887...93G},
(7) -- {\it {\it {\it {\it Gaia}}}} DR3 distance in kpc followed by an uncertainty in parentheses,
(8) -- the SED slope at wavelengths $\lambda \ge 10 \mu$m defined as d $[\log (\lambda$ F$_\lambda)$]/d $[\log \lambda$]),  
(9) $-$ observatory acronym{\bf , where our spectra were taken}: SPM -- Observatorio Astron\'omico Nacional San Pedro Martir (Mexico), MCD - McDonald Observatory (TX, USA), CFHT -- Canada-France-Hawai'i Telescope (HI, USA), MO -- Mount Orzale (Italy), IAO -- Indian Astronomical Observatory (Hanle, India), LO -- Lick Observatory (CA, USA).\\
$^a$ -- for UCAC4\,452--018064 two emission components of the H$\alpha$ line above the photospheric wings were measured separately and summed up.
}
\end{table*}

\begin{table*}
\caption{TShAO photometric data}\label{tab3}
    \setlength{\tabcolsep}{2pt}
\begin{center}
\begin{tabular}{llccccllcccc}
\hline\noalign{\smallskip} 
 UCAC4 & Alt. ID & HJD & $V$     &  $B-V$ & $V-R$ & UCAC4 & Alt. ID & HJD & $V$ &  $B-V$ & $V-R$ \\
       &         &     & mag	  &  mag   & mag   &       &         &     & mag &  mag	  &   mag \\ 
\noalign{\smallskip}\hline\noalign{\smallskip}
\multicolumn{1}{c}{(1)}&
\multicolumn{1}{c}{(2)}&
\multicolumn{1}{c}{(3)}&
\multicolumn{1}{c}{(4)}&
\multicolumn{1}{c}{(5)}&
\multicolumn{1}{c}{(6)}&
\multicolumn{1}{c}{(1)}&
\multicolumn{1}{c}{(2)}&
\multicolumn{1}{c}{(3)}&
\multicolumn{1}{c}{(4)}&
\multicolumn{1}{c}{(5)}&
\multicolumn{1}{c}{(6)}\\
\noalign{\smallskip}\hline\noalign{\smallskip}

615--019793  & HBHA\,3215--11 & 2457384.329	& 12.25	& 0.43	& 0.19  & 708--091648 &	HBHA\,5204--04   &	2457681.197 & 12.71 & 0.61	&  0.38  \\
             &		         & 2457430.235 & 12.27 & 0.45  & 0.17  &             &		            &  2457689.043 & 12.72 & 0.64  &  0.37  \\
             &		         & 2457452.115 & 12.28 & 0.42  & 0.19  &             &		            &  2457698.062 & 12.76 & 0.64  &  0.37  \\
712--078173	 & HBHA\,5202--07 &	2459138.135 & 13.17	& 1.26	& 0.35	&             &		          	 &	2457722.144 & 12.90	& 0.60	&  0.36  \\
             &                & 2459139.216 & 13.24 & 1.26  & 0.37  &             &		            &  2458739.312 & 12.85 & 0.59  &  0.35  \\
             &		         & 2459164.120 & 13.19 & 1.30  & 0.36  &             &		            &  2458783.258 & 12.87 & 0.58  &  0.36  \\
             &		         & 2459205.023 & 13.20 & 1.21  & 0.36  &			 &		            &  2458783.290 & 12.88 & 0.57  &  0.36  \\
757--021798  & HBHA\,6211--04 &	2457384.186	& 14.38	& 0.99	& 0.52  &             &		          	 &	2460854.370 & 12.82	& 0.61	&  0.39  \\
753--023300	 &IRAS\,01571+6018&	2456864.292 & 15.43 & 0.86	& 0.75	&			  &		      	     &	2460915.146 & 12.92	& 0.58	&  0.36  \\
             &                & 2456870.281 & 15.48 & 0.88  & 0.75  &			 &		            &  2460922.229 & 12.97 & 0.58  &  0.38  \\
668--034110	 & HBHA\,4215--01 &	2457720.381 & 12.57	& 0.35  & 0.27	& 403--115685 &	IRAS\,19156--0935&  2460585.156 & 12.01 & 0.43	&  0.30  \\
             &                & 2457771.242 & 12.57 & 0.34  & 0.29  &			 &                  &  2460602.066 & 11.79 & 0.65  &  0.64  \\
             &                & 2457792.070 & 12.57 & 0.35  & 0.28  &             &                  &  2460609.132 & 11.67 & 0.61  &  0.45  \\
             &                & 2457798.103 & 12.56 & 0.34  & 0.28  &	         &                  &  2460615.059 & 11.71 & 0.60  &  0.55  \\
             &                & 2457818.178 & 12.57 & 0.35  & 0.29  &             &                  &  2461200.392 & 11.69 & 0.56  &  0.58  \\
             &                & 2457841.086 & 12.60 & 0.34  & 0.29  &			 &                  &  2461279.193 & 11.87 & 0.39  &  0.29  \\
             &                & 2457841.086 & 12.55 & 0.35  & 0.28  & 579--016291 & HBHA\,2715--26   &  2461278.383 & 12.62 & 0.39  &  0.16  \\
\noalign{\smallskip}\hline
\end{tabular}
\end{center}
\tablecomments{Column information: 
(1) -- ID from the {\it UCAC4} catalog,
(2) -- Alternative ID,
(3) -- Heliocentric Julian Date of the observational mid-time,
(4) -- $V$--band brightness,
(5--6) -- $B-V$ and $V-R$ color-indices.\\
}
\end{table*}

\begin{figure}
    \centering
    \includegraphics[width=0.9\linewidth,height=0.88\textheight,keepaspectratio]{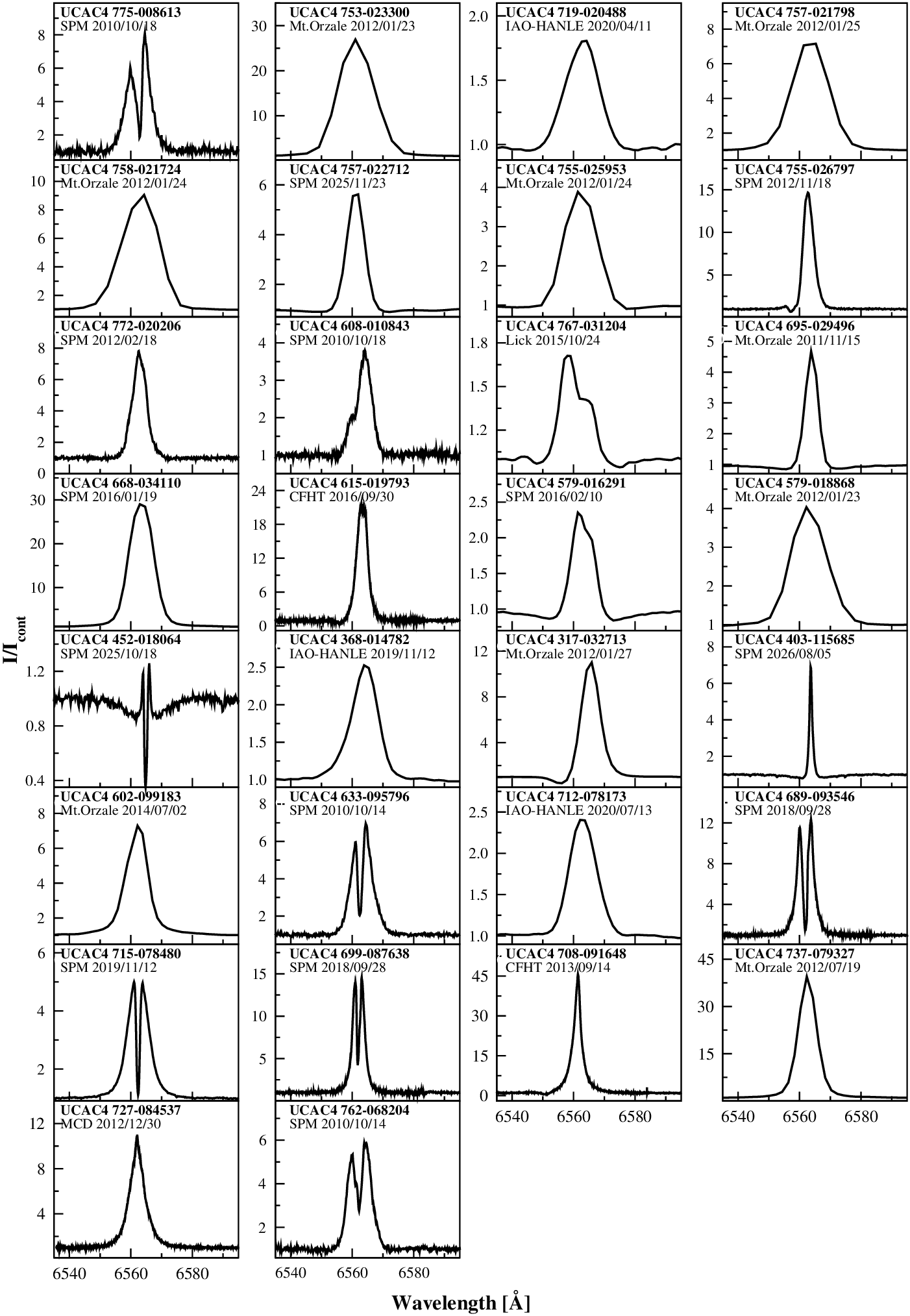}
    \caption{The objects' continuum-normalized H$\alpha$ line profiles. Objects' IDs, telescope used, and observing dates are shown in each panel. The wavelength scale is heliocentric. }
    \label{fig2}
\end{figure}

\begin{figure*}[t]
    \centering
    \includegraphics[width=0.85\linewidth,height=0.57\textheight,keepaspectratio]{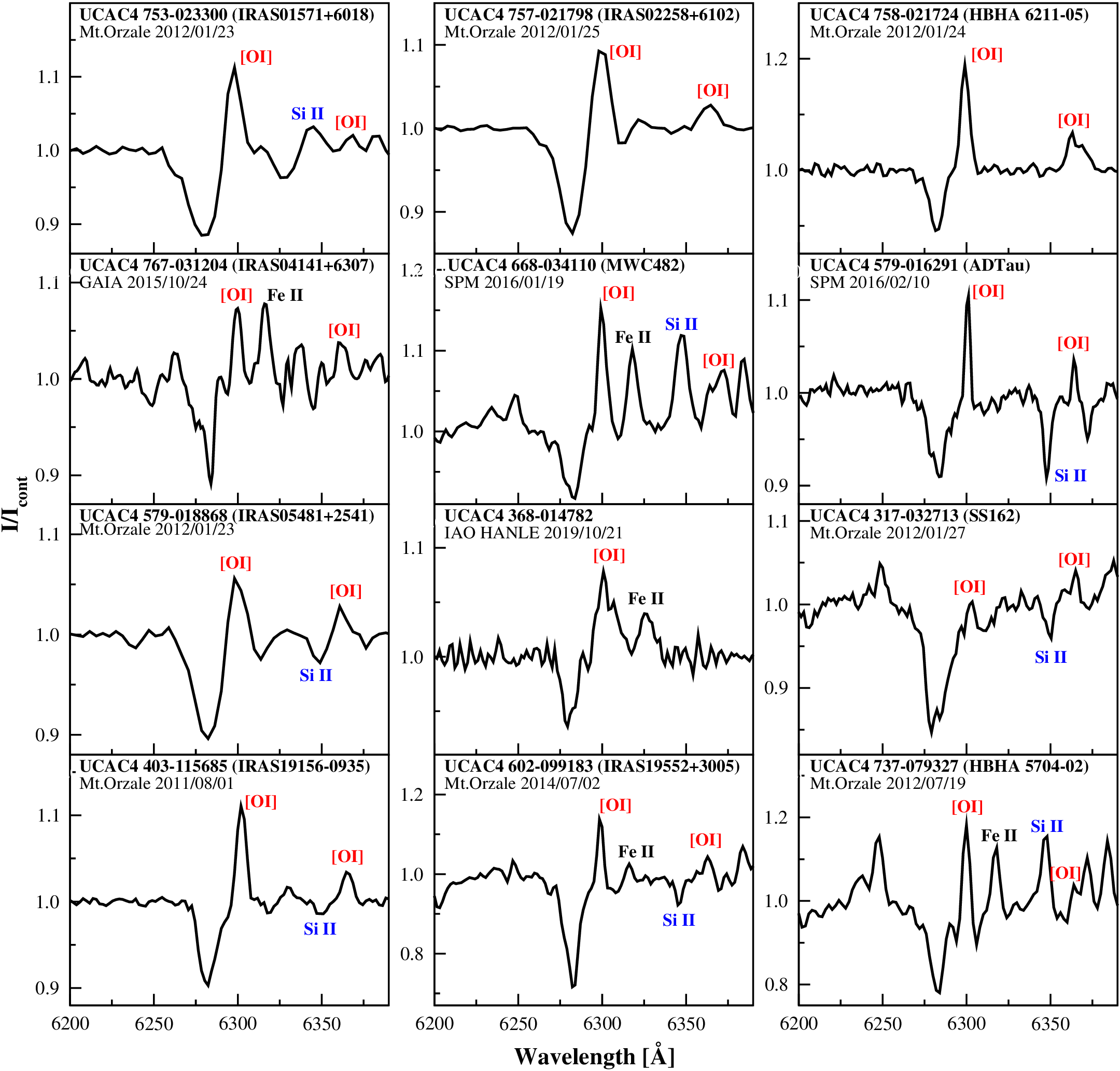}\\
    \includegraphics[width=0.6\linewidth,height=0.32\textheight,keepaspectratio]{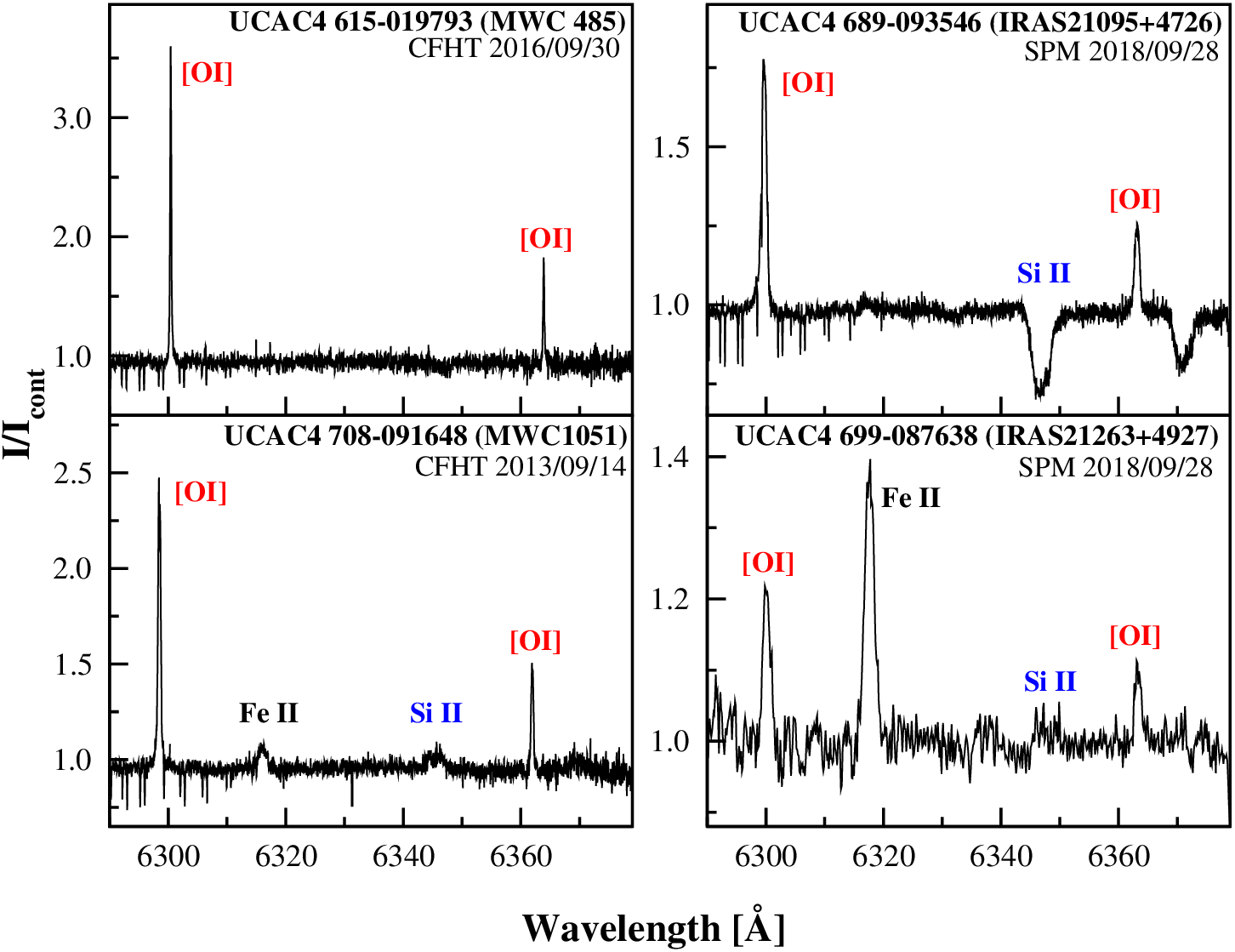}
    \caption{Parts of the objects' in which forbidden [O {\sc i}] lines were detected. The upper 12 spectra are low-resolution ($R = 800-1500$), while the lower 4 spectra are high-resolution ($R \sim 65,000$). Objects' IDs, telescopes used, and observing dates are shown in each panel. Identified emission and absorption lines are marked. The wavelength scale is heliocentric, the intensity is normalized to the local continuum.}
    \label{fig3}
\end{figure*}

\begin{figure*}[!htb]
     \includegraphics[width=0.95\linewidth]{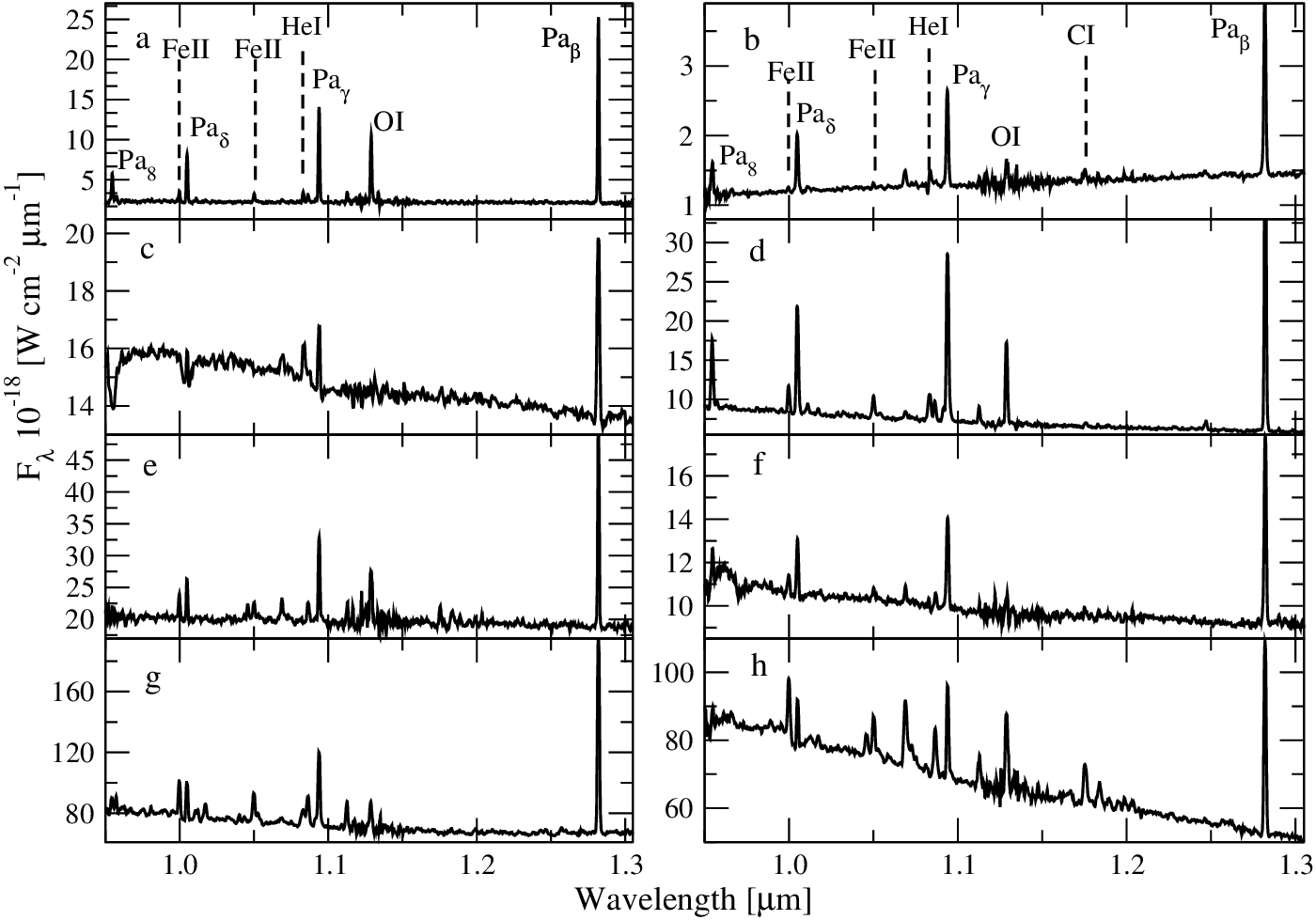}
    \caption{Parts of the low-resolution near-IR flux-calibrated spectra taken at the Lick Observatory. Panels show spectra of the following objects {\bf(observing dates are shown in parentheses after each object): a) IRAS\,01571+6018 (07/21/2013), b) IRAS\,02258+6102 (07/21/2013), c) IRAS\,04265+4845 (11/11/2011), d) IRAS\,04577+4328 (10/07/2017), e) IRAS\,19552+3005 (06/15/2014), f) IRAS\,20098+3622 (09/10/2013), g) IRAS\,21263+4927 (06/22/2010), and h) AS\,466 (07/03/2012).} Fluxes are shown in units of $10^{-18}$ W\,m$^{-2}$\,$\mu$m$^{-1}$, wavelengths are shown in microns. The identified emission lines are marked in the top panels and can be easily recognized in all other panels.
     }
    \label{fig4}
\end{figure*}

\begin{figure*}[!htb]
     \includegraphics[width=0.95\linewidth]{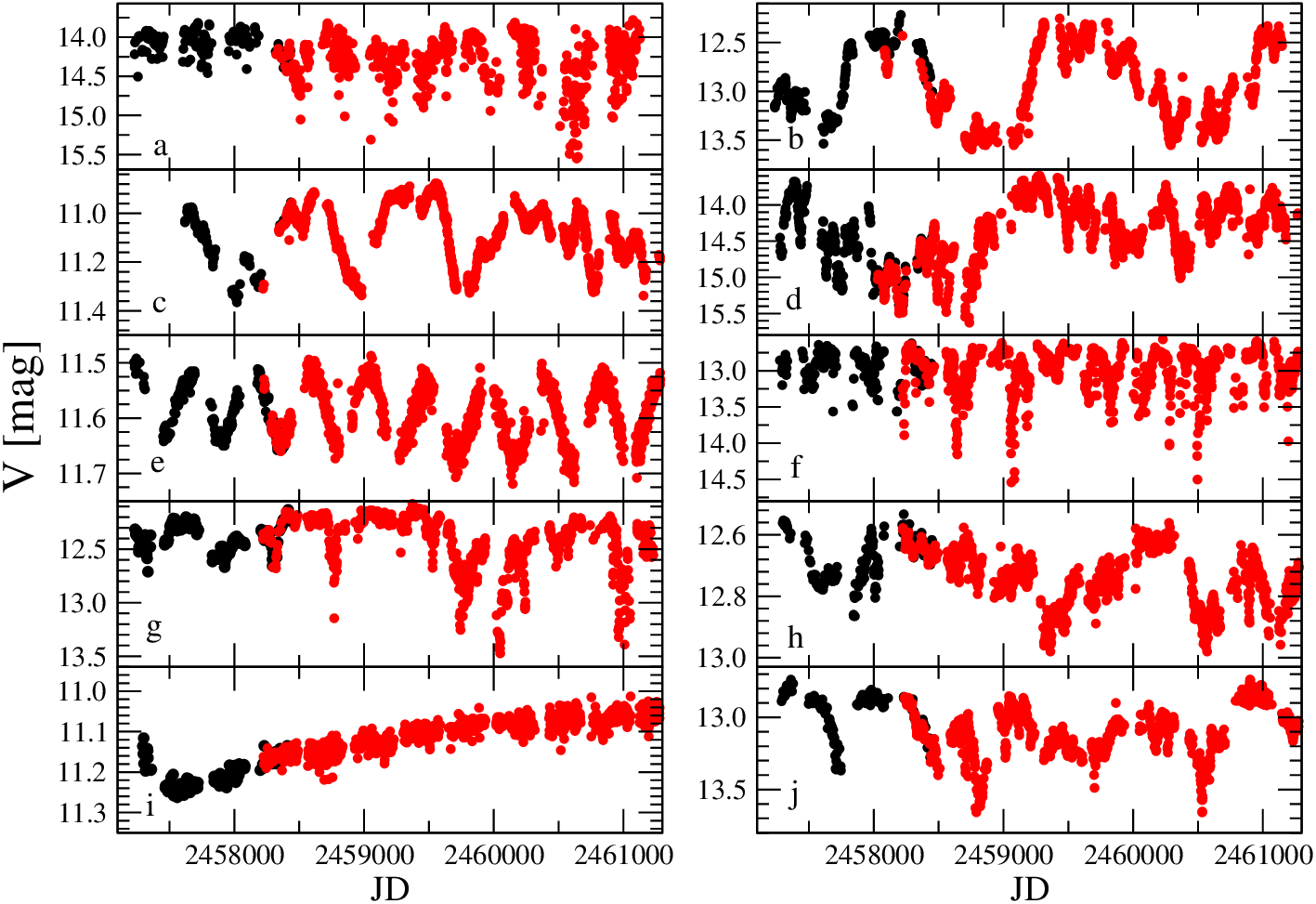}
    \caption{ASAS SN light curves of the most variable objects. $V$--band data are shown by black circles, $g$--band data are shifted to match the brightness level in the $V$--band and shown by red circles. Uncertainties of the shown data do not exceed 0.05 mag.
    The objects' IDs and $g$--band data shifts are listed below. \\
    Panels: a) IRAS\,04141+6307 ($g - 0.50$), b) AD\,Tau ($g - 0.00$), c) IRAS\,06237+0020 ($g - 0.42$), d) UCAC4\,368--014782 ($g - 0.40$), e) IRAS\,19156$-$0935 ($g - 0.37$), f) KT\,Cyg ($g - 0.14$), g) IRAS\,21095+4726 ($g - 0.33$), h) IRAS\,21263+4927 ($g - 0.40$), i) AS 466 ($g - 0.57$), j) IRAS\,22541+6205 ($g - 0.48$).
    }
    \label{fig5}
\end{figure*}

\begin{figure*}[t]
    \centering
    \includegraphics[width=0.48\linewidth]{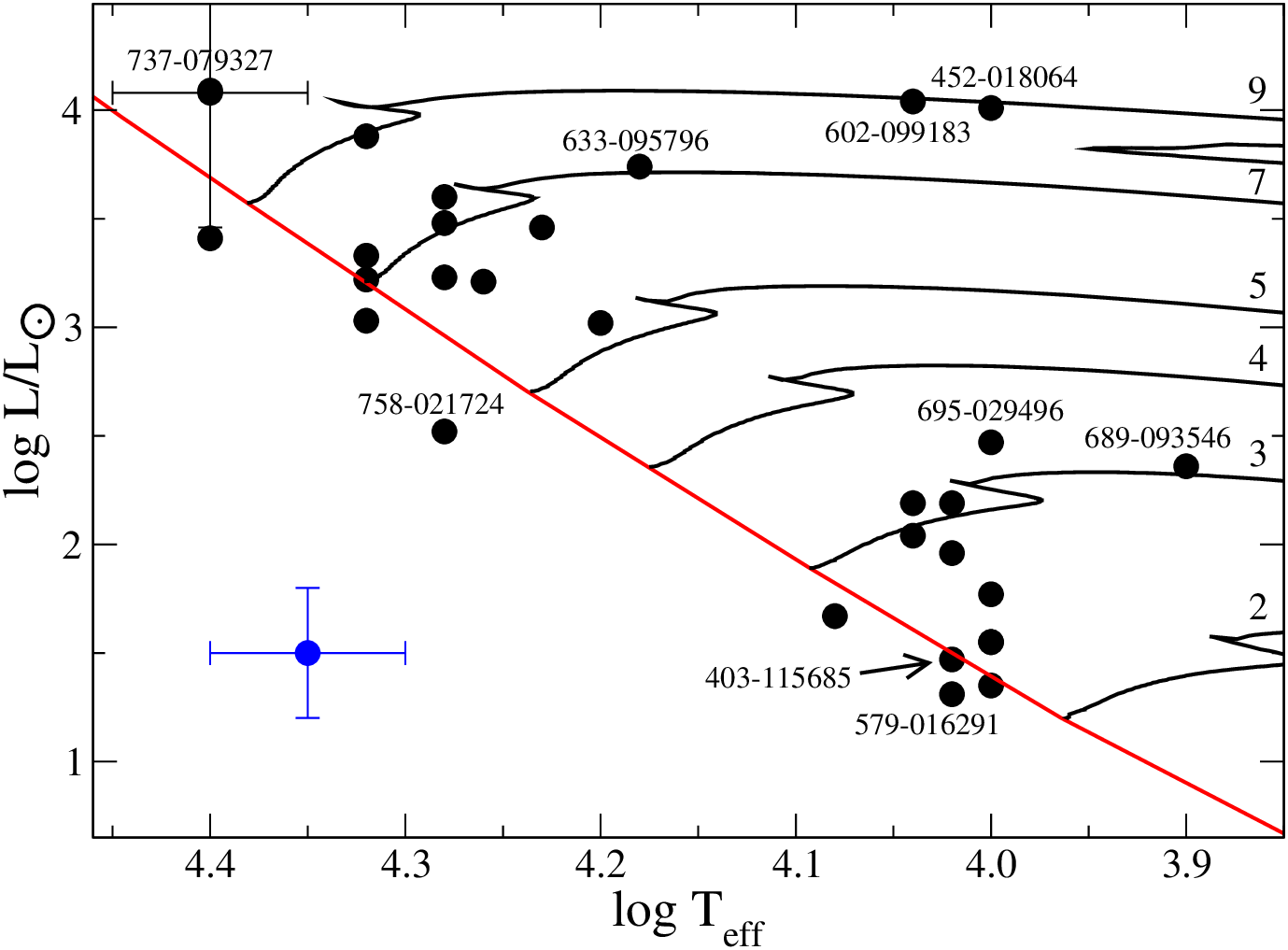}\hfill
    \includegraphics[width=0.50\linewidth]{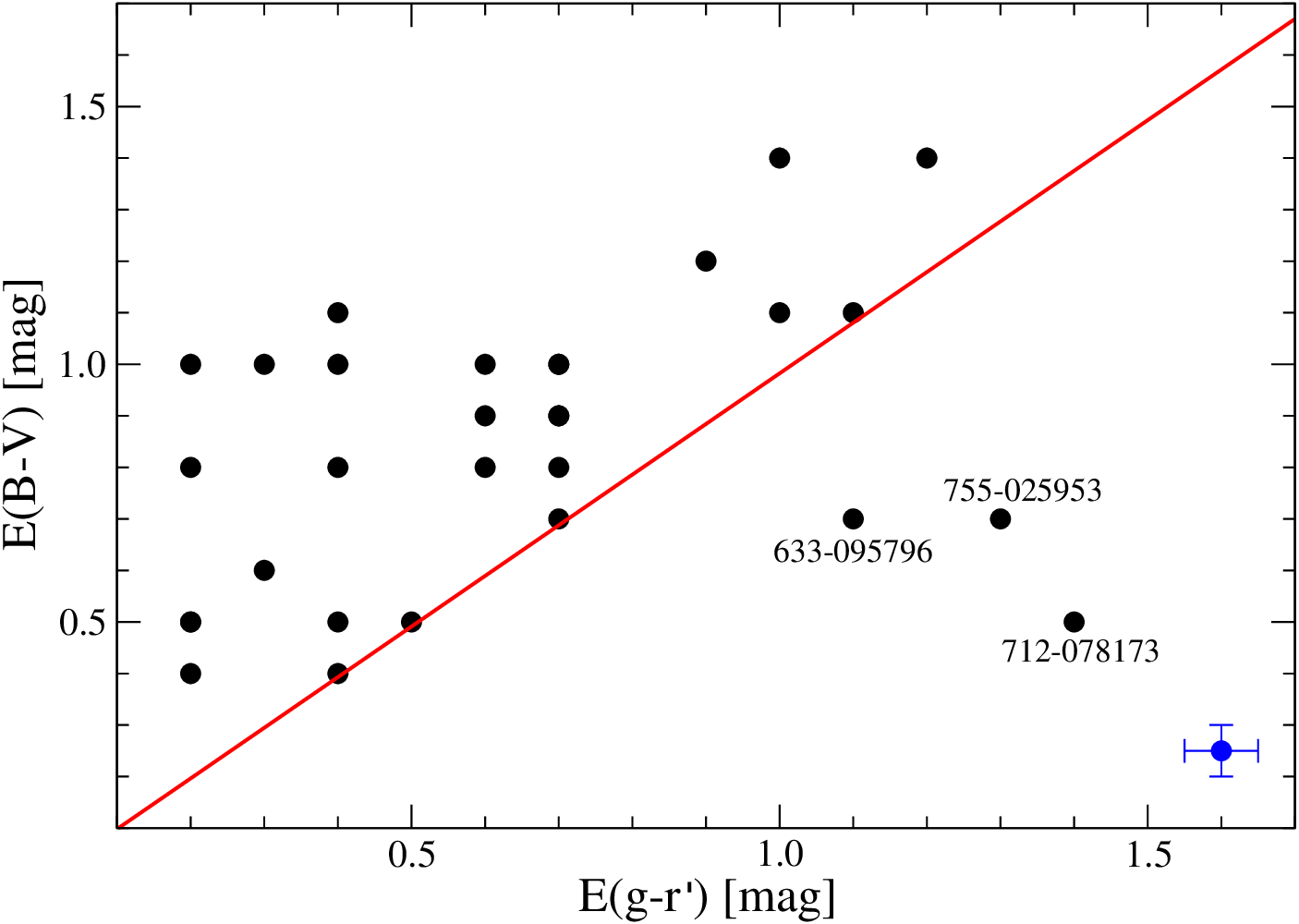}
    \caption{
    {\bf Left panel:} HRD for the calculated objects' parameters shown in Table\,\ref{tab2} along with their uncertainties. UCAC4 IDs are indicated at the positions of selected objects, whose parameters are discussed in Sect.\,\ref{properties}. Numbers near the evolutionary tracks for single rotating stars taken from \citet{2012A&A...537A.146E} indicate initial masses in solar units. The red solid line shows the zero-age main-sequence. The blue cross shows typical uncertainties of the fundamental parameters for all objects except for UCAC4\,737--079327, which has the largest uncertainty of the luminosity due to a large error in the {\it Gaia} DR3 distance (see Table\,\ref{tab2}).\\
    {\bf Right panel:} Comparison of color-excesses E$(B-V)$ derived from the SED fitting and E$(g-r^{\prime})$ from the interstellar extinction maps compiled by \citet{2019ApJ...887...93G}. The red solid line indicates the color-excess relationship E$(B-V) = 1.05 \times$ E$(g-r^{\prime})$ from \citet{2016ApJ...821...78S}. {\bf The blue cross shows typical uncertainties of the plotted color-excesses (see Sect.\,\ref{search} for the explanation)}. Three objects with E$(g-r^{\prime})$ noticeably larger than E$(B-V)$ are individually discussed in Sect.\,\ref{properties}.
        }
    \label{fig6}
\end{figure*}

\section{Results}

All 30 selected objects show emission lines and IR excesses that demonstrate different shapes, although most IR fluxes fall noticeably toward longer wavelengths in the mid-IR spectral range ($\lambda\lambda \sim 5-25 \mu$m). The latter feature is explained by a lack of distant cold CS dust that may be due to the compactness of the dusty shell or a lack of UV photons from the underlying star that warm the dust most efficiently. We characterize the excess shape by assessing a SED slope defined as $d[\log(\lambda F_{\lambda})]/d[\log \lambda]$ longward of $\lambda \sim 5 \mu$m (see column ``Slope'' in Table\,\ref{tab2} and Fig.\,\ref{fig1}). The objects with the Slope of $\leq 0.7$ may be young (pre-main-sequence) stars with large CS envelopes/disks left from the star-formation phase, while those with steeper slopes may have CS dust formed during later stages of their evolution, such as mass-transfer in binary systems or a strong mass-loss at the AGB stage.

The H$\alpha$ emission line profiles for all sample objects are shown in Fig.\,\ref{fig2}, parts of the optical spectra with different resolutions for objects with detected forbidden [O {\sc i}] 6300 \& 6363 \AA\, emission lines are shown in Fig.\,\ref{fig3}, while examples of the near-IR low-resolution spectra near $\lambda = 1 \mu$m are shown in Fig.\,\ref{fig4}. The objects are marked by their UCAC4 IDs in the Figures, while some alternative IDs are listed in Table\,\ref{tab1} and Figures\,\ref{fig3}--\ref{fig6}.

To assess the objects' brightness variations, light curves in the $V$-- and $g$--band were collected from the ASAS SN survey \citep{2017PASP..129j4502K} between the time of its start ($\sim$ 2015) and August--September 2026. 
Data taken in one day were averaged. The averaged data with dispersion of $\ge 0.05$ mag were removed from the final datasets. The resulting light curves for objects with the strongest brightness variations are shown in Fig.\,\ref{fig5}. The average brightness in the $V$--band for objects with small variations was used in the luminosity calculations along with the spectroscopic information (see Sect.\,\ref{properties}) and the total extinction (interstellar + CS) determined from the SED fitting (see Fig.\,\ref{fig1}). The highest brightness was used for the most variable objects, assuming that the dimming is due to variable obscuration by CS matter. A search for periodicity in the objects' brightness was performed using the NASA Exoplanet Archive periodogram service\footnote{https://exoplanetarchive.ipac.caltech.edu/cgi-bin/Pgram/nph-pgram}.

We also compared the E$(B-V)$ color-excesses, which represent the total reddening, derived from the SED fitting with {\bf the purely interstellar color-excesses E$(g-r^{\prime})$} calculated from the 3D Bayestar reddening maps by \citet{2019ApJ...887...93G}. The results are shown in Table\,\ref{tab2} and Fig.\,\ref{fig6}. In most cases, the total reddening was found to be larger than the interstellar one, which qualitatively indicates that the SED fits are reasonable. Three objects for which reversed results were found are discussed individually in Sect.\,\ref{properties}.

In the next Section, we review the previously published information available for the objects, describe their properties, which we deduce from the observational data taken by us and found in the archives and databases, and suggest our own assessment of the objects' nature and evolutionary status.\\

\subsection{Properties of individual objects} \label{properties}

\subsubsection{Newly found emission-line objects} \label{new}

UCAC4\,775--008613. It shows a double-peaked H$\alpha$ line in emission and a relatively steep decrease in the IR excess longward of $\lambda \sim 2 \mu$m. Our SPM spectrum shows emission-line components within the interstellar Na {\sc i} D-lines and the He {\sc i} 5876 \AA\, line in absorption. The latter is consistent with a T$_{\rm eff}$ of $\sim 10,000$ K derived from the SED fitting. A larger E$(B-V)$ color-excess compared to that expected from the purely interstellar extinction (see Table\,\ref{tab2}) suggests the presence of some CS extinction in the line of sight. As no star-forming region has been found near the object, it is a good candidate to the FS\,CMa group with the B[e] phenomenon.\\

UCAC4\,772--020206 = 2MASS J03094366+6419040. This is a 13 magnitude star-like object, which was classified as a galaxy LEDA\,2797261 for its extended appearance \citep{2003A&A...412...45P} along with its association with the IR source IRAS\,03054+6407. The images from various surveys, such as DSS and 2MASS, reveal the presence of two fainter star-like objects on both sides of the star. The star's positions in UCAC4, 2MASS, WISE, and AKARI (at 9 \& 18 $\mu$m) coincide with each other, while the IRAS source position is $\sim 17\arcsec$ South of the star. Our SPM spectrum shows a strong single-peaked H$\alpha$ emission line, a weak H$\beta$ emission inside broad absorption wings, and several Fe {\sc ii} emission lines. The WISE images at 11 \& 22 $\mu$m (W3 and W4 bands) exhibit an arc-shaped structure centered on the star, which seems to be an expanding dusty shell towards the South-East. A nearly flat IR excess along with the presence of the arc probably indicates that it is a young star.\\

UCAC4\,767--031204 = IRAS\,04141+6307. An isolated object that shows the H$\alpha$ emission line while other Balmer lines and Paschen lines are seen in absorption. The ASAS SN light curve exhibits strong brightness variations ($g$--mag between $\sim$14.5 and $\sim$16.5 mag, see Fig.\,\ref{fig5}a). Periodogram analysis of the light curve revealed the highest peak in the power spectrum at 298 days. The peak is broad and the phase curve folded with this period displays a significant scatter, which is unlikely to occur as a result of eclipses. The nearly flat IR excess between $\lambda \sim 1$ and 22~$\mu$m may point to its pre-main-sequence evolutionary status, but the IR flux has not been detected at longer wavelengths. Multicolor photometry and higher-resolution spectroscopy are needed to uncover the reasons for the variability and to check the presence of the B[e] phenomenon.\\

UCAC4\,368--014782 = 2MASS\,J06301654--1629375. The only published result for this object is the inclusion in a catalog of R\.CrB--type star candidates by \citet{2024PASP..136h4201K}, where it was suggested to be at minimum brightness with a mostly featureless near-IR spectrum that showed the He{\sc i} 10830 \AA\, line in absorption.
Our low-resolution spectra taken near maximum brightness (see Fig.\,\ref{fig5}d) show the H$\alpha$ line and the Ca {\sc ii} IR triplet in emission. A weak forbidden line of [O {\sc i}] at 6300 \AA\, was marginally detected. The presence of the G--band at 4300 \AA\, and numerous weak absorption lines suggest that a cool star dominates the blue part of the spectrum, but no other molecular features were detected. The shape of the IR excess implies a lack of cold dust. A BP/RP Gaia spectrum taken in a wavelength range between 336 and 1020 nm can be reproduced with a model spectrum for T$_{\rm eff} \sim 5,500$ K  \citep{2003IAUS..210P.A20C} and E$(B-V) \sim 0.4$ mag. It is unlikely a T~Tau star, as it is not located in a star-forming region.
We suggest that it is a binary system with the B[e] phenomenon, although higher-resolution spectroscopy is needed to refine its nature.\\

\subsubsection{Previously suggested young star candidates}

UCAC4\,719--020488 = IRAS\,02162+5326. The object has only been mentioned as a young star candidate by \citet{2014PASJ...66...17T} based on AKARI far-IR data without description of its properties. The images taken by the WISE mission show the presence of a stream or a jet at 11 and 22 $\mu$m. Our low-resolution HCT spectrum shows the only emission in the H$\alpha$ line, whereas the higher members of the Balmer series are strong and seen in absorption, which is consistent with our T$_{\rm eff} = 11,000$ K. Our low-resolution near-IR spectrum shows only weak Pa$\beta$ and Br$\gamma$ emission lines. There is no nebulosity around the object on optical images, while there is a molecular cloud at 18$\arcsec$ SE from it that is not aligned with a recently discovered jet  \citep{2016A&A...594A..28P}. Given the HRD position within the main-sequence (at $\sim$ 3 M$\odot$, see Table\,\ref{tab2} and Fig.\,\ref{fig5}), the object may be a young star near the end of its pre-main-sequence stage, when the protostellar CS disk is getting dispersed.\\

UCAC4\,757-021798 = IRAS\,02258+6102. It is located in the direction of the star-forming region W3 but slightly further away (by $\sim200$ pc) and does not exhibit CO line emission \citep{2003A&A...399.1083K}. The optical and near-IR spectrum is rich with emission lines, and overall properties are similar to those of UCAC4\,753--023300 (see Sect.\,\ref{late_stages}). The SED slope in mid-IR (see Table\,\ref{tab2}) is not well constrained because the fluxes at $\lambda \ge 10 \mu$m are below 1 Jy.
The object shows the B[e] phenomenon and is more likely an FS\,CMa system rather than a pre-main-sequence star. We note here that \citet{2023ApJ...942...22C} included it in a list of pre-main-sequence stars and detected emission lines of the hydrogen Brackett series, but more evolved hot stars can also have them formed by other processes than accretion.\\

UCAC4\,758-021724 = HBHA\,6211-–05. Although this object was included in a sample of classical Be star candidates by \citet{2016A&A...591A.140G} and listed as a Be star in SIMBAD, it obviously has CS dust and most likely a pre-main-sequence star mentioned in a number of recent papers \citep[e.g.,][]{2023AJ....166..183V}. However, the latter authors derived a lower T$_{\rm eff}$ of $\sim 8,500$ K compared to our estimate of 19,000 K. Our low-resolution spectrum shows the He {\sc i} 5876 \AA\, line in emission that points to an early B--type of the underlying star. In addition, the Balmer jump strength seen in the {\it Gaia} BP/RP spectrum is also consistent with the higher T$_{\rm eff}$. The {\it Gaia} DR3 distance is consistent with that of the W3 region and, within the uncertainties of its luminosity and T$_{\rm eff}$, places it on HRD very close to the zero-age main-sequence (see Fig.\,\ref{fig6}). The latter results further support the status of UCAC4\,758-021724 as a young stellar object.\\

UCAC4\,757--022712 = IPHAS J023950.95+611829.1 and UCAC4\,755--025953 = HBHA 6212--36.
Similarly to the previous object, these two were classified as early Ae stars by \citet{2016A&A...591A.140G}. Our T$_{\rm eff}$ are slightly higher (see Table\,\ref{tab2}), but a large and steadily declining IR flux is consistent with the presence of mostly hot/warm CS dust. They are definitely not classical Be stars. They are both located within the W3/W4/W5 star-forming clouds at distances consistent with that adopted for the complex.
UCAC4\,757-022712 was considered a young stellar object by \citet{2019AJ....157..112P}, while UCAC4\,755-025953 was included in a list of young stars by \citet{2008ApJ...688.1142K} and \citet{2023AJ....166...97L}. Our low-resolution spectra of both objects show emission components within broad photospheric absorptions of the H$\beta$ lines and an easily recognizable Mg {\sc ii} 4482 \AA\, absorption line, which is consistent with the derived T$_{\rm eff}$. Although the 3D Bayestar map shows very different color-excesses E$(g-r^{\prime})$ for the objects, which are separated by $\sim 1^{\circ}$ (see Fig.\,\ref{fig6}), the equivalent width of the diffuse interstellar band at $\lambda\,5780$ \AA\, in their spectra is almost the same, thus suggesting nearly the same interstellar extinction toward them. The Bayestar map shows a very narrow region of higher extinction near UCAC4\,755--025953, and the resulting color-excess may be unreliable if its true distance is even a little different from the adopted {\it Gaia} DR3 value.
Only the IR flux of UCAC4\,757--022712 was detected at $\lambda \leq 22 \mu$m (the Herschel mission measured a flux of $434\pm30$ mJy at $\lambda = 70 \mu$m). These results do not support the objects' young nature, because of the lack of cold CS dust. Higher-quality spectra are needed for these objects to see whether the B[e] phenomenon is present.
\\

UCAC4\,755-026797 = IRAS\,02551+6042. The last object in our list from the W3/W4/W5 region. Initially classified as a B4e star by \citet{2016A&A...591A.140G} and considered a young star by \citet{2008ApJ...688.1142K,2023AJ....166..183V,2023AJ....166...97L}. However, \citet{2023AJ....165...51R} quote $\log$ T$_{\rm eff} = 3.8$, which is significantly lower than what is expected for a B4-type star. Our SED and the optical spectrum analysis support the higher T$_{\rm eff}$. The Balmer line profiles show a prominent blue-shifted absorption component (P\,Cyg--type), which is indicative of a strong stellar wind. A nearly flat IR SED is consistent with a young star status.\\

UCAC4\,608--010843 = IRAS\,03519+3123. The object is classified as an AGB candidate in SIMBAD, while several studies considered it a young star \citep[e.g.,][]{2021MNRAS.503.3232P} in an open cluster in Perseus star-forming clouds. \citet{2022A&A...668A...4F} list a T$_{\rm eff}$ of $\sim$6,350 K and a higher than solar metallicity. Our spectra show the H$\alpha$ and H$\beta$ lines in emission, consistent with a T$_{\rm eff} \sim 12,000$ K from the SED analysis. The IR excess is weak and declines toward longer wavelengths. If it does belong to a cluster, its far-IR flux may evaporate before the near-IR flux does due to the UV radiation of nearby stars. Therefore, it can be a pre-main-sequence object near the end of this evolutionary stage.\\

UCAC4\,695--029496 = IRAS\,04265+4845. It was only recently identified as a young star \citep{2022ApJ...930...39V} that belongs to a small cluster \citep{2023AJ....166..183V}. Our visual and near-IR low-resolution spectra show not only hydrogen emission lines, such as H$\alpha$, Pa$\beta$, and Br$\gamma$, but also He {\sc i} lines at 1.083 and 2.058 $\mu$m in emission. This is unusual for a late--B type star, whose T$_{\rm eff}$ derived by \citet{2022ApJ...930...39V} and us are nearly the same and also consistent with the Balmer jump seen in the {\it Gaia} BP/RP spectrum, and may indicate the presence of a hotter component in the system. The IR excess decreasing toward loner wavelengths along with a non-detection of the IR fluxes at long-wavelength passbands of IRAS and AKARI and the HRD location near the terminal-age main-sequence (see Fig.\,\ref{fig6}) make the object's young status dubious.\\

UCAC4\,668-034110 = IRAS\,04577+4328 = MWC\,482. First classified as a R\,CrB type object \citet{2012A&A...539A..51T}, later suggested as a young star \citep{2022ApJ...930...39V}, but appeared again in a R\,CrB objects list \citep{2024PASP..136h4201K}. Our data show the presence of the B[e] phenomenon along with a strong emission-line spectrum and a steady IR flux decline longward of $\sim 5 \mu$m. The emission in the lines of the Balmer (H$\alpha$ and H$\beta$), Paschen, and Brackett (Br$\gamma$) series along with that in He {\sc i} lines (5876 \AA\, 1.083 and 2.058 $\mu$m) suggest a high T$_{\rm eff}$ (see Table\,\ref{tab2}) corroborated by \citet{2022ApJ...930...39V}. We do not see compelling evidence of the pre-main-sequence status of MWC\,482 and suggest it to be a candidate to the FS\,CMa objects group.\\

UCAC4\,615--019793 = IRAS\,05111+3244 = MWC\,485. There have been three different suggestions about its nature: a spiral galaxy \citep{1994A&AS..104..529T}, a transitional young stellar object \citep{1999A&A...352..228M}, and a possible merger in an evolved system \citep{2024A&A...689A.234D}. The IRAS and WISE images show nebulous structures very close to the star. Our high and low-resolution spectra show hydrogen Balmer series (H$\alpha$ through H$\gamma$), many Paschen series lines, and Br$\gamma$ in emission, as well as the He {\sc i} IR emission lines at 1.083 and 2.058 $\mu$m. There also narrow forbidden [O {\sc i}] lines at 6300 and 6363 \AA\, (see Fig.\,\ref{fig3}). This line content is more consistent with a higher T$_{\rm eff}$ derived by us (see Table\,\ref{tab2}) than with T$_{\rm eff} = 10,000$ K  derived by \citet{2022ApJ...930...39V}. The rising IR flux longward of $\lambda \sim 10 \mu$m favors the pre-main sequence status, although it does not reject the merger scenario assuming that the surrounding dusty clouds are consequences of this event.\\

UCAC4\,317--032713 = SS\,162. Discovered as an emission-line star in an objective prism survey \citep{1977ApJS...33..459S}, the object was included in a sample of Herbig Be star candidates based on a {\it Gaia} BP/RP spectrum \citep{2023A&A...679A..71S}. Our low-resolution spectra (although with a higher resolution than that from {\it Gaia}) show H$\alpha$ through H$\gamma$ lines in emission with P\,Cyg type profiles that indicate a strong stellar wind along with emission lines of Fe {\sc ii}. Lines of neutral helium are seen in absorption. The detection of forbidden lines in the visual spectral region is unclear (see Fig.\,\ref{fig3}). The described features of the optical spectrum are consistent with a T$_{\rm eff}$ between 15,000 and 20,000 K, which is in agreement with T$_{\rm eff}$ = 17,353\,K (with no uncertainty) given in \citet{2023A&A...679A..71S}. However, the distance listed in Table A.1 of the latter paper ($7641^{+2}_{-4}$\,K) seems to have an overestimated precision and contradicts the value calculated from the {\it Gaia} DR3 parallax \citep{2021AJ....161..147B} and shown in Table\,\ref{tab2}. The luminosity calculated from the average $V$--band brightness taken from the ASAS SN data ($V = 13.25\pm0.03$ mag), the interstellar extinction (A$_{V} = 2.5\pm0.3$ mag), the {\it Gaia} DR3 distance, and an appropriate bolometric correction (BC$_{V} = -1.5\pm0.2$ mag) puts the object near the zero-age main-sequence. Given a clustering of nearly a dozen stars at roughly the same {\it Gaia} DR3 distance, we suggest a pre-main-sequence status for SS\,162.\\

UCAC4\,689--093546 = IRAS\,21095+4726. Discovered as an emission-line star by \citet{1956BOTT....2n..19G}, it was suggested to be a young star in a Bok globule B\,361 \citep{1983PASJ...35..299O}. Its ASAS SN light curve led to a suggestion that it was a weak R\,CrB star candidate \citep{2019MNRAS.483.4470S} but classified as an irregular variable by these authors (see Fig.\,\ref{fig5}g).
The object also appears in several recent lists of Herbig Ae/Be star candidates \citep[e.g.,][]{2022ApJ...930...39V}. \citet{2026ApJ..1005...41W} observed it at $\lambda = 1.3$ mm, but have detected no signal pointing to a lack of cold CS dust around it. Our medium and high-resolution optical spectra show double-peaked Balmer lines and Fe {\sc ii} lines in emission, He {\sc i} lines in pure absorption, and forbidden emission lines of [O {\sc i}] at 6300 and 6363 \AA. The Paschen series hydrogen lines show weak emission components. The He {\sc i} 4471 \AA\, absorption line that is noticeably weaker than the Mg {\sc ii} 4482 \AA\, line is consistent with a T$_{\rm eff} \le 10,000$ K. Overall, the observed properties, including the position on HRD beyond the main-sequence (see Fig.\,\ref{fig6}), are more consistent with classification as an FS\,CMa type object. \citet{2024A&A...689A.234D} consider IRAS\,21095+4726 a potential merger in a binary system, but currently available data do not allow us to test this hypothesis.\\

UCAC4\,699--087638 = IRAS\,21263+4927. \citet{2023A&A...679A..71S} included it in their sample of Herbig Be stars, while
\citet{2024PASP..136h4201K} suggested that it is an R\,CrB candidate star near its brightness maximum. Our high-resolution spectra taken at SPM and CFHT are similar to those of IRAS\,21095+4726 but with stronger Fe {\sc ii} lines and hydrogen Paschen series lines in emission. The photospheric He {\sc i} 4471 \AA\, line is noticeably stronger than that of the Mg {\sc ii} 4482 \AA\, line suggesting a T$_{\rm eff}$ between 15,000 and 20,000~K which is in agreement with the SED fitting. The shape of the IR excess implies a lack of cold dust. IRAS\,21263+4927 was also included in the list of potential mergers by \citet{2024A&A...689A.234D}. However, similarly to IRAS\,21095+4726, this hypothesis currently cannot be tested. Therefore, we tentatively  classify it as an FS\,CMa type object.\\

UCAC4\,762--068204 = IRAS\,22541+6205. Initially detected by the IRAS mission and the Hamburg-Bergedorf survey, it was suggested to be a young star candidate due to its location near the Cepheus OB3 association area \citep{2001Ap&SS.275..441M}. \citet{2012ApJ...750..125A} considered it a class {\sc II} young stellar object based on the Spitzer MIPS colors in the 3.5 -- 8 $\mu$m wavelength range. It was also selected as a member of an open cluster UBC\_178 recognized using {\it Gaia} DR3 astrometry \citep{2022ApJS..262....7H}. At the same time, the {\it Gaia} DR3 distance to the object ($\sim 900$ pc) is larger than that to Cep OB3 (720 pc). Also, although the IR flux is nearly flat between $\lambda \sim 1$ and 25 $\mu$m ($\sim2$ Jy), the source was not detected in the long-wavelength bands of IRAS ($\lambda = 60~ \& ~100 \mu$m) and AKARI ($\lambda \ge 18 \mu$m), while the flux at $\lambda = 70~\mu$m detected by Herschel is much weaker ($0.81\pm0.05$ Jy) implying a lack of cold CS dust. The WISE images do not show any extended emission. Therefore, the evolutionary status of IRAS\,22541+6205 as a young star remains dubious. Periodogram analysis of the ASAS SN $g$--band light curve (see Fig.\,\ref{fig5}j) indicates regular variations with a period of 915 days and an amplitude of $\sim$0.7 mag. Since only a few cycles have been observed, it is important to keep monitoring the object and take a better spectrum to refine its spectral classification.

\subsubsection{Late evolutionary state candidates} \label{late_stages}

UCAC4\,753--023300 = IRAS\,01571+6018. Based on IR color-indices, it was considered to be either a young Planetary Nebula candidate \citep{2009A&A...502..113V} or a R\,CrB type system \citep{2024PASP..136h4201K}, but both suggestions were not confirmed in closer studies. \citet{2003A&A...399.1083K} detected the CO line emission in the direction of the object, but no H\,{\sc ii} region was found there \citep{2021ApJS..253...23A}. A steep decrease of the mid-IR flux towards longer wavelengths (Fig.\,\ref{fig1}) and a strong H$\alpha$ emission (see Table\,\ref{tab2} and Fig.\,\ref{fig2}) along with the presence of the B[e] phenomenon (Fig.\,\ref{fig3}) make the properties of UCAC4\,753--023300 similar to those of hot members of the FS\,CMa group \citep{2023Galax..11...36M}.\\

UCAC4\,579--016291 = AD\,Tau. SIMBAD lists its spectral type as B2 and assigns to a category of ``long period variables''. The H$\alpha$ line in emission was discovered in the course of the Hamburg-Bergedorf survey. Our low-resolution spectrum suggests a lower T$_{\rm eff}$ ($\sim 10,500$ K, see Table\,\ref{tab2}) due to a stronger Mg {\sc ii} 4482 \AA\, photospheric line compared to that of the He {\sc i} 4471 \AA\, line.
At the same time, the forbidden [O {\sc i}] emission lines at 6300 and 6363 \AA\ (see Fig.\,\ref{fig3}) are clearly present as signs of the B[e] phenomenon. Weak diffuse interstellar bands and nearly equal brightness in the $V$-- and $g$--bands ASAS SN data  also support our T$_{\rm eff}$ estimate. The latter also show a strong brightness variability of $\Delta$(V, g) $\sim 1.3$ mag with two cycles of $\sim 300$ days \citep{2020ApJS..247...44A} and $\sim$1,500 days easily visible in the data (see Fig.\,\ref{fig5}b). The longer cycle was also detected by the INTEGRAL telescope \citep{2012A&A...548A..79A}.
The IR-excess peaks at $\lambda \sim 10~\mu$m and shows a steep flux decrease toward longer wavelengths (see Fig.\,\ref{fig1}) suggesting a lack of cold CS dust. In general, the described properties along with a low luminosity (see Table\,\ref{tab2} and Fig.\,\ref{fig6}) can be interpreted as signs of a binary system, perhaps of the FS\,CMa type.\\

UCAC4\,403--115685 = IRAS\,19156$-$0935. This object appeared in a list of candidates for OH/IR maser emission, but the maser lines were not detected \citep{1992A&A...254..133L}.
It was also suspected to belong to the hydrogen-deficient and carbon-rich supergiant R\,CrB stars \citep{2012A&A...539A..51T}.
The SED shows a strong IR excess, whose integrated energy exceeds that in the shorter-wavelength region where the star dominates the emergent flux. This is a sign of a strong, mostly gray CS extinction, similar to those of the proto-planetary nebulae Red Rectangle and IRAS\,07080+0605 \citep{2022ApJ...932...36K}. If true, its HRD position (see Fig.\,\ref{fig6}) needs to be corrected toward a larger luminosity. 
Similarly to IRAS\,07080+0605, UCAC4\,403--115685 shows no surrounding nebulosity that can be explained by a star or a binary system at an early proto-planetary stage. A search for periodicity of the spectral-line positions would be needed to check for binarity of this object. The ASAS SN light curve (see Fig.\,\ref{fig5}e) shows a regular cyclic process with a period of 446 days and an amplitude of 0.17 mag.\\

UCAC4\,452-018064 = IRAS\,06237+0020. It was classified as an M2/3 type star \citep{McConnell1993}, although our spectrum, which has a relatively low signal-to-noise ratio, is typical for a late--B type star with prominent hydrogen lines and very few metallic lines. This spectrum is reminiscent of binary post-AGB systems with a strong metallic depletion due to their locking in CS dusty particles \citep{2019A&A...629A..49O}. Its luminosity L $\sim 10^4$ L$_{\odot}$ (see Fig.\,\ref{fig6}) is consistent with that of an intermediate-mass star that can move toward higher T$_{\rm eff}$ fast enough to exhibit a large change of the spectral type within a few decades. The IR excess shape is also similar to that of post-AGB stars. {\bf The ASAS light curve shows a cyclic behavior with a period $\sim 950$ days and an amplitude of $\sim 0.4$ mag in both $V$ and $g$ bands (see Fig.\,\ref{fig5}c), but smaller brightness variations are also noticeable.}\\

UCAC4\,633-095796 = IRAS\,20098+3622 = KT\,Cyg. The object was discovered as a variable star by \citet{1930MiSon..17....1H}. Spectral type B5e was assigned to it by \citet{1960ApJ...131..632H}. It was included in a catalog of Be stars by \citet{1982IAUS...98..261J} and also in a {\it {\it {\it {\it Gaia}}}} DR3 catalog of eclipsing binaries \citep{2023A&A...674A..16M}. The IR excess was noticed by \citet{2004NewA....9..509Z}. Our optical and near-IR spectra show the presence of hydrogen emission lines (H$\alpha$, Paschen series, and Br$\gamma$) and weak forbidden oxygen lines at 6300 and 6363 \AA. The IR excess declines longward of $\lambda \sim 3 \mu$m. Our T$_{\rm eff}$ from the SED fitting (15,000 K) is consistent with the assignment of the spectral type by \citet{1960ApJ...131..632H}. As seen in Fig.\,\ref{fig6}, the calculated interstellar reddening from the 3D maps is larger than the total reddening derived from the SED fit. However, an earlier study by \citet{1980A&AS...42..251N}, although based on a small number of stars, showed an interstellar reddening consistent with our result. The discrepancy may be due to different regions used for the reddening calculations and is beyond the scope of our paper.
The ASAS SN light curve (see Fig.\,\ref{fig5}f) demonstrates brightness minima of various depths (0.5 -- 1.8 mag), whose occurrence are described by a dominant peak in a Lomb-Scargle periodogram at 237 days. This periodicity is not strictly eclipsing and exhibits a significant scatter near the minima. It may be a combination of eclipses by a companion and a variable CS matter distribution. In general, the observed properties suggest that KT\,Cyg belongs to the binary group of FS\,CMa type objects.\\

\subsubsection{Objects with no previously assigned or uncertain status}

UCAC4\,579--018868 = IRAS\,05481+2541. Discovered as an emission line star by \citet{1956BOTT....2n..19G} and later confirmed by \citet{1989AbaOB..66...19C} and \citet{1999A&AS..134..255K} at low spectral resolution, this object has not been studied in depth. A cometary nebula is visible on optical and IR images of the object. Our low-resolution spectra contain moderately strong H$\alpha$ line and weak [O {\sc i}] 6300 and 6363 \AA\, emission lines, while higher Balmer series lines are seen in pure absorption. The slow decline of the IR flux with wavelength along with the presence of the nebula and a location near the zero-age main-sequence (see Table\,\ref{tab2}) suggest a transitional young star status.\\

UCAC4\,602--099183 = IRAS\,19552+3005. The object was reported as an emission-line star by \citet{1999A&AS..134..255K} and detected by IRAS without further studies. Our low-resolution optical and near-IR spectra show many emission lines of hydrogen (H$\alpha$, Paschen series, and Br$\gamma$) and iron (Fe {\sc ii}) as well as weak forbidden oxygen lines at 6300 and 6363 \AA. The IR excess has a strong decline longward of $\lambda \sim 5~\mu$m which is similar to those of evolved objects with the B[e] phenomenon (e.g., FS\,CMa type objects). The {\it Gaia} DR3 distance seems to be too large and leads to a luminosity over 10$^4$ L$_{\odot}$ (see Fig.\,\ref{fig6}), but the interstellar reddening derived from the SED fitting and the Bayestar extinction map \citep{2019ApJ...887...93G} are consistent with each other. A higher-resolution spectrum is needed to constrain the object's fundamental parameters.\\

UCAC4\,712--078173 = HBHA\,5202--07. The H$\alpha$ line was detected in the Hamburg-Bergedorf survey. It was suggested to be a possible member of one of the young clusters within $\sim$1 kpc from the Sun by \citet{2019AJ....158..122K} and \citet{2022ApJS..262....7H}. The IR fluxes were only detected by 2MASS and WISE. Our low-resolution optical spectra show the presence of a strong H$\alpha$ (see Fig.\,\ref{fig2}) and a barely recognizable H$\beta$ line in emission along with many weak absorption lines, including the CH band at $\lambda = 4300$ \AA, indicative of a late-type spectrum. The near-IR spectrum shows the He {\sc I} 10830 \AA\, emission line, but the hydrogen lines are not clearly recognizable. Similarly to UCAC4\,368--014782 (see Sect.\,\ref{new}), a {\it Gaia} BP/RP spectrum can be reproduced with a cool-star model spectrum for T$_{\rm eff} \sim 5,500$ K \citep{2003IAUS..210P.A20C} and E$(B-V) \sim 0.5$ mag. The WISE image at $\lambda = 22 \mu$m ($W4$--band) shows a narrow annular emission around the star that may point to its early evolutionary stage. Comparison of the object's area image at various wavelengths suggests that it is projected on a dark cloud, which may be a star-forming region. Otherwise, its properties can be interpreted by a binary system model, such as the FS\,CMa type object MWC\,645 \citep{2022ApJ...936..129N}, in which the hot companion is only revealed by the emission-line presence. A higher resolution spectrum is needed to resolve this controversy. The discrepancy between our color-excess derived from SED fitting and that calculated from the Bayestar map may be due to the uncertainties of the adopted {\it Gaia} DR3 distance possibly caused by the complex content of the object.
\\

UCAC4\,715-078480 = AS\,466. Discovered in the course of the Mount Wilson search for emission-line stars \citep{1950ApJ...112...72M}, the presence of H$\alpha$ emission was confirmed by \citet{1999A&AS..134..255K}, while the IR excess was noticed by \citet{2005MNRAS.363.1111C}. Our medium- and high-resolution optical spectra contain numerous double-peaked hydrogen and Fe {\sc ii} emission lines, while the forbidden [O {\sc i}] lines are barely noticeable. The near-IR spectrum contains a strong IR Ca {\sc ii} triplet and the O {\sc i} 8446 \AA\, line in emission. Weak Paschen lines and Br$\gamma$ in emission were also detected. The ASAS SN light curve exhibits smooth behavior with a weak dimming at the end of 2015 followed by a slow brightening (see Fig.\,\ref{fig5}i). The shape of the IR excess rules out its pre-main-sequence status, while unusually strong IR Ca {\sc ii} triplet is atypical in classical Be stars. We suggest that AS\,466 is most likely a binary system with a CS gaseous disk, which is too dense for forbidden line formation, but also with CS dust present near the object.\\

UCAC4\,708--091648 = IRAS\,21525+5113. The H$\alpha$ line emission-line star was discovered in the Mount Wilson survey \citep[MWC\,1051,][]{1949ApJ...110..387M}, confirmed by the Hamburg-Bergedorf survey, while its IR excess was noticed by \citet{2016MNRAS.463.1162C}.  Using an automated algorithm, \citet{2020A&A...638A.103C} selected it as a likely central star of a Planetary Nebula, while \citet{2022ApJ...930...39V} included it in their list of Herbig Ae/Be stars. Our data show that it has a strong emission-line spectrum, exhibits the B[e] phenomenon, and a declining IR flux longward of $\lambda \sim 5 \mu$m. These features are similar to those of objects in the FS\,CMa group.\\

UCAC4\,737-079327 = HBHA\,5704--02. This emission-line star was discovered in the Hamburg-Bergedorf survey and selected as a Planetary Nebula \citep{2009A&A...502..113V} and a R\,CrB candidate \citep{2024PASP..136h4201K}, but no detailed study was presented in these papers. Our low-resolution spectrum shows Balmer emission lines (H$\alpha$ through H$\delta$), He {\sc i} (5876 \& 6678 \AA), O {\sc i} IR triplet and Fe {\sc ii} lines. The B[e] phenomenon is represented by the [O {\sc i}] 6300 \AA\, line, while the 6363 \AA\, line is too weak. The combination of a strong emission-line spectrum, the IR excess shape, and the fundamental parameters (see Fig.\,\ref{fig6}) matches the observed properties of the FS\,CMa group members \citep[cf.][]{2007ApJ...667..497M}.\\

UCAC4\,727-084537 = AS\,487. This emission-line star was suggested to have a high luminosity by \citet{1964LS....C03....0H} and was later included in a catalog of hot luminous stars \citep{2003AJ....125.2531R}. It was also mentioned in statistical studies of the local spiral structure \citep{2021A&A...645L...8X} and open clusters \citep{2018A&A...615A..49C}, but its individual properties were not addressed. Our high-resolution spectra show double-peaked emission lines of hydrogen, Fe {\sc ii}, and O {\sc i} (7772--7775 \AA\, \& 8446 \AA), while He {\sc i} lines are observed in absorption. The absorption lines are wide, indicating a high rotational velocity, which, along with a relatively low luminosity (see Table\,\ref{tab2}) suggests that the star is a dwarf. The IR excess is relatively small and may be mostly due to the CS gas. A ``wine-bottle'' shape of the Balmer line profiles along with the broad photospheric lines may imply that the CS disk is oriented nearly edge-on, and the central peak in the line profiles may form in a stream of material transferred to a secondary companion.\\

\section{Discussion} \label{discussion}

Although most of the objects in our sample were discovered decades ago, they were typically studied using limited information, while their nature and/or evolutionary status were assigned from statistical studies of large groups with similar properties. Our selection was based mostly on the near-IR color-indices, but our study used all currently available information from the literature in as broad a spectral range as possible, as well as our own photometric and spectroscopic observations. As a result of this broader view, previous classification of 7 young stellar objects was confirmed or corroborated, and that of 2 such objects was considered dubious, 17 objects were confidently reclassified, and classification for 4 newly found emission-line stars was suggested.

At the same time, some long-standing difficulties in classification of stars with IR excesses were encountered despite the amount of information collected. In particular, separating pre-main-sequence Herbig Ae/Be stars from more evolved objects with the B[e] phenomenon is still a challenging task. The former may or may not exhibit forbidden lines, while it is a major defining criterion for the latter. Also, Herbig stars exhibit mostly flat IR excesses at wavelengths $\lambda \leq 100 \mu$m and lose the near-IR excess earlier that far-IR excess, while objects with the B[e] phenomenon, such as FS\,CMa objects, show a noticeable decline of the IR excess at wavelengths $\lambda \geq 5-10 \mu$m \citep{2007ApJ...667..497M}.

When early-type pre-main-sequence objects were recognized \citep{1960ApJS....4..337H}, one of the defining criteria was the presence of a visual nebulosity around them. Later, isolated pre-main-sequence stars that showed IR excesses and linear polarization of optical light were recognized \citep[e.g.,][]{1991Ap&SS.186..283G}, but no nebulosities were found near them. Also, it is harder to detect a nebulosity near distant objects as well as IR fluxes at mid- and far-IR wavelengths, which is another defining feature of young stats. Therefore, assigning a young status to objects that luck either cold dust or do not belong to star-forming regions is still problematic.

On the other hand, comparison of the derived color-excesses from the SED fitting and those expected at the objects' distances from the interstellar extinction maps by \citet{2019ApJ...887...93G} became an additional check of the results reliability. For objects that demonstrate larger amounts of the observed reddening measured by color-excesses than the interstellar reddening at their distances, it is a sign of CS reddening. In cases of the opposite (more interstellar than total reddening), it may point to a {\it Gaia} distance error that was found to be a problem for objects with CS envelopes \citep[e.g.,][]{2022A&A...667A..74A}.

\section{Conclusions} \label{conclusions}

Our sample includes 14 young stellar object candidates (mostly Herbig Be stars), 5 objects that were suggested to be evolved stars, 7 objects with no previously assigned status, and 4 newly found emission-line stars (see Sect.\,\ref{properties}).
Five objects were found to be located in or near the W3/W4/W5 star-formation complex and suggested to be pre-main-sequence stars based on the location and large IR fluxes. Our analysis supports this suggestion only for two of them (UCAC4\,758--021724 and UCAC4\,755--026797). The other 3 objects (UCAC4\,757-021798, UCAC4\,757--022712, and UCAC4\,755--025953) lack evidence of the presence of distant cold CS dust. The former exhibits forbidden emission lines and is more likely an FS\,CMa object, while the latter two still need far-IR flux measurements to get their status constrained.

We found the properties of other 4 objects (UCAC4\,719--020488, UCAC4\,608--010843, UCAC4\,615--019793, and UCAC4\,317--032713) consistent with the status of young stars. A jet-like structure was revealed in IR images of the former object. We also conclude on the pre-main-sequence status of UCAC4\,579--018868, whose IR image contains a cometary nebula around it.
At the same time, we found no compelling arguments to support the young status for 8 other objects that were suggested to be young stars in the literature. They are either located outside of star-forming regions or have a well-defined decline of the IR flux longward of $\lambda \sim 5 \mu$m.

A subgroup of 12 objects: UCAC4\,775--008613, UCAC4\,757--021798 (mentioned above), UCAC4\,753--023300, UCAC4\,668--034110, UCAC4\,602--099183, UCAC4\,689--093546, UCAC4\,699--087638, UCAC4\,737--079327, UCAC4\,633--095796, UCAC4\,712--078173, UCAC4\,708--091648, and possibly UCAC4\,368--014782 were found to be strong candidates to the subgroup of objects with the B[e] phenomenon known as FS\,CMa type objects. They may represent a fainter extension of this group of evolved mass-transferring possible binaries or merger products with large gaseous and compact dusty disks \citep[e.g.,][]{2007ApJ...667..497M,2023Galax..11...36M,2024A&A...689A.234D}. Additionally, UCAC4\,715--078480 may also belong to the FS\,CMa group, but it may be surrounded by a dense gaseous disk that suppresses forbidden-line formation in it.

Three objects (UCAC4\,579–016291, UCAC4\,403--115685, and UCAC4\,452--018064) exhibit long-term quasi-periodic cycles of the optical brightness variations with periods of several years that are typical of stars at late evolutionary stages, such the Proto-Planetary stage. One of them (UCAC4\,403--115685) has an extremely strong IR excess, implying a large extinction of the stellar radiation in an optically-thick dusty envelope. This property is similar to that of a remarkable Proto-Planetary Nebula Red Rectangle but with no visible nebula.
Another one (UCAC4\,579–016291) may still be an FS\,CMa type binary system at a more advanced evolutionary stage.

In order to constrain the still uncertain classification of 11 objects (see descriptions in Sect.\,\ref{properties}), at least high-resolution spectroscopy and far-IR photometry would be very helpful. In the future, we are planning to search for similar objects in the Southern hemisphere of the sky and fill the gaps in the information about those reported in this paper.

\section{Acknowledgements}
This work was supported by the Ministry of Science and Higher Education of the Republic of Kazakhstan (Grant No. AP23489575). 
The work is carried out within the framework of the Project No. BR31714731 "Fundamental research in observational and theoretical astrophysics", financed by the Ministry of Science and Higher Education of the Republic of Kazakhstan.
S.Z. acknowledges PAPIIT grants IN102120 and IN119323.
This research has made use of the SIMBAD database, operated at CDS, Strasbourg, France; SAO/NASA ADS. This paper is partly based on observations obtained at the 2.7\.m Harlan J. Smith telescope of the McDonald Observatory (Mt. Locke, TX, USA), the 2.1\,m telescope of the Observatorio Astron\'omico Nacional San Pedro Martir (Baja California, M\'exico), the 3\,m telescope of the Lick Observatory (California, USA), the 1.5\,m telescope of the Bologna Observatory (Mt. Orzale, Italy), the 2\,m Chandra Himalayan Telescope (Hanle, India), and a 1\,m of the Tien-Shan Astronomical Observatory (Almaty, Kazakhstan). 

The research is also based on observations obtained at the Canada-France-Hawai'i Telescope (CFHT) which is operated by the National Research Council of Canada, the Institut National des Sciences de l'Univers of the Centre National de la Recherche Scientifique of France, and the University of Hawai'i. CFHT is located on Maunakea on Hawai'i Island, a mountain of considerable cultural, natural, and ecological significance. Maunakea is a sacred site to Native Hawaiians, also known as K\~anaka '\=Oiwi. Quality observations are made possible by relentless effort of the entire staff at Canada-France-Hawai'i Telescope.

This work is partly based on results from the European Space Agency (ESA) space mission {\it Gaia}. {\it Gaia} data are being processed by the {\it Gaia} Data Processing and Analysis Consortium (DPAC). Funding for the DPAC is provided by national institutions, in particular the institutions participating in the {\it Gaia} MultiLateral Agreement (MLA). The {\it Gaia} mission website is https://www.cosmos.esa.int/Gaia. The {\it Gaia} archive website is https://archives.esac.esa.int/Gaia.

\bibliographystyle{aasjournal}
\bibliography{main}

\end{document}